\documentclass[twocolumn,amsmath,amssymb,floats,superscriptaddress,nofootinbib,10pt]{revtex4-1}

\pdfoutput=1

\AtBeginDocument{%
    \newwrite\bibnotes
    \def\bibnotesext{Notes.bib}
    \immediate\openout\bibnotes=\jobname\bibnotesext
    \immediate\write\bibnotes{@CONTROL{%
    apsrev41Control,author="08",editor="1",pages="1",title="0",year="1"}}
     \if@filesw
     \immediate\write\@auxout{\string\citation{apsrev41Control}}%
    \fi
}%

\usepackage{graphicx}
\usepackage{dcolumn}
\usepackage{bm}
\usepackage{revsymb}
\usepackage[usenames]{color}
\usepackage{color}
\usepackage{physics}
\usepackage{amsfonts}
\usepackage[section]{placeins} 
 
\usepackage{amsmath}
\usepackage{ulem}
\usepackage[usenames]{color}
\usepackage{amsfonts}
\usepackage{graphicx}
\usepackage{dcolumn}
\usepackage{bm}
\usepackage{revsymb}
\usepackage[usenames]{color}
\usepackage{graphicx}

\usepackage{color}
\usepackage{physics}
\usepackage{amsmath}
\usepackage{amssymb,bbm}
\usepackage{lipsum}  
\usepackage{float}

\usepackage{subcaption}
\usepackage{tikz}
\usetikzlibrary{arrows.meta}   
\usetikzlibrary{calc}          

\usepackage[version=4]{mhchem}

\usepackage{xcolor}

\newcommand{\bbv}{{\mathbf{v}}}
\newcommand{\bbw}{{\mathbf{w}}}

\newcommand{\me}{{\mathbf{e}}}
\newcommand{\mep}{{\mathbf{e}'}}

\newcommand{\bea}{\begin{eqnarray}}
\newcommand{\eea}{\end{eqnarray}}

\newcommand{\kb} {{k_\text{b}}}
\newcommand{\mur} {{\mu_\text{r}}}
\newcommand{\bird} {{\text{B}}}
\newcommand{\air} {{\text{A}}}
\newcommand{\self} {\text{self}}

\newcommand{\flap} {\text{flap}}

\newcommand{\graze} {\text{graze}}

\newcommand{\inv} {{\text{I}}}

\usepackage{bbm}

\definecolor{nblue}{RGB}{28,130,185}
\definecolor{cgreen}{RGB}{76,153,0}
\definecolor{myorange}{RGB}{245,156,74}

\usepackage{hyperref}
\hypersetup{
  colorlinks=true,
  citecolor=magenta,
  urlcolor=-myorange
}

\usepackage{xcolor}
\definecolor{ogreen} {RGB}{71,191,145}
\newcommand{\fl}[1]{\textcolor{black}{#1}}

\usepackage{hyperref}
\hypersetup{
  colorlinks=true,
  citecolor=magenta,
  urlcolor=-myorange
}

\newcommand{\quotes}[1]{``#1''}

\begin{document}

\title{A microscopically reversible kinetic theory of flocking}

\author{Ruben Lier}
\email{r.lier@uva.nl}
\affiliation{Institute for Theoretical Physics, University of Amsterdam, 1090 GL Amsterdam, The Netherlands}
\affiliation{Dutch Institute for Emergent Phenomena (DIEP), University of Amsterdam, 1090 GL Amsterdam, The Netherlands}
\affiliation{Institute for Advanced Study, University of Amsterdam, Oude Turfmarkt 147, 1012 GC Amsterdam, The Netherlands}

\begin{abstract}
We formulate a kinetic theory of two species of hard spheres undergoing reactive collisions that convert chemical energy into kinetic energy. \textcolor{black}{The model describes an active species interacting with a passive background, labeled as \quotes{birds} and \quotes{air} respectively, with the reactive collisions representing of self-propulsion. Microscopic reversibility of the reactive dynamics is imposed, and a chemostat is introduced to drive the system out of equilibrium. When the chemostat is sufficiently strong and one restricts to grazing interspecies collisions, we find that the bird momentum damping coefficient can change sign, giving rise to a flocking transition.} \end{abstract}

\maketitle

\tableofcontents
\section{Introduction}
The Toner-Tu equations are one of the earliest and most widely studied continuum descriptions of active matter \cite{toner1995longrange,toner1998flocks,TONER2005170} and they are given by
\begin{subequations} \label{eq:tonertuequations}
    \begin{align}  
\begin{split}
\partial_t \mathbf{v}_0  
&+ \lambda_1 (\mathbf{v}_0  \cdot \vec{\nabla}) \mathbf{v}_0 
+ \lambda_2 (\vec{\nabla} \cdot \mathbf{v}_0 ) \mathbf{v}_0 
+ \lambda_3 \vec{\nabla}(|\mathbf{v}_0 |^2)  \\
&=  - \alpha \mathbf{v}_0  - \beta |\mathbf{v}_0 |^2 \mathbf{v}_0  
- \vec{\nabla} P_1 
- \mathbf{v}_0  (\mathbf{v}_0  \cdot \vec{\nabla} P_2) 
 \\
& + D_1 \vec{\nabla} (\vec{\nabla} \cdot \mathbf{v}_0 ) + D_T \nabla^2 \mathbf{v}_0  
+ D_2 (\mathbf{v}_0  \cdot \vec{\nabla})^2 \mathbf{v}_0 , \end{split}
 \\
\partial_t n_{\bird } &+ \vec{\nabla} \cdot (\mathbf{v}_0  n_{\bird }) = 0, 
\end{align}
\end{subequations}
where $n_{\bird }$ is a bird density, $\mathbf{v}_0 $ is the bird velocity, $P_{1,2}$ are functions that depend on $|\mathbf{v}_0 |$ and $n_{\bird }$. A key feature of \eqref{eq:tonertuequations} is the \quotes{flocking transition} that it can display when $\alpha < 0 $ and $\beta  > 0 $. 
\newline 
The precise nature of the flocking transition and its universal scaling properties have been subject to debate for more than a decade \cite{toner2012reanalysis,PhysRevE.110.054108,PhysRevLett.132.268302,chen2025inconvenienttruthflocks,PhysRevE.109.044126,PhysRevLett.123.218001,PhysRevE.107.024611}. These disagreements notwithstanding, it is understood that the reason bird flocks described by \eqref{eq:tonertuequations} can display a flocking transition whereas an ordinary fluid described by the Navier-Stokes equation cannot, lies in two characteristics of bird flocks that separate it from an ordinary fluid:
\begin{enumerate}
    \item  \label{item:first} 
    Birds are embedded in a \textit{medium of air}. Therefore, a fluid equation like \eqref{eq:tonertuequations} that describes only the flow of birds is not constrained  by Galilean invariance or momentum conservation. This is the reason that several terms in \eqref{eq:tonertuequations}, including the $\alpha-$ and $\beta$-terms, are allowed to exist. 
    \item  \label{item:second} Birds are \textit{active}, which means they can burn fuel at the particle level and thereby circumvent constraints that would otherwise follow from the second law of thermodynamics. In this case, the constraint is that the coefficient $\alpha$ should always be positive.  
\end{enumerate}
It follows from the above-mentioned points that it is the coefficient $\alpha$ in \eqref{eq:tonertuequations} that lies at the heart of the flocking transition.
\newline 
To better appreciate why bird flocks are fundamentally different from ordinary fluids, it is helpful to look at the microscopic origin of flocking, which could stem from different underlying processes \cite{PhysRevLett.112.148102,musacchio2025flockingsecondorderphasetransition,Agranov_2024,proesmans2025quantifyingdissipationflockingdynamics,PhysRevLett.119.058002,PhysRevLett.96.104302,PhysRevLett.110.055702,PhysRevLett.112.075701,PhysRevLett.130.148202,PhysRevX.14.031008,Baconnier_2025,PhysRevLett.111.268302,Grossman_2008,PhysRevLett.105.098001}.
The earliest and most widely studied microscopic model for flocking is the Vicsek model \cite{Vicsek_1995,Czirók_1997,Ginelli_2016,PhysRevLett.92.025702,PhysRevE.77.046113,PhysRevLett.105.168103,PhysRevE.92.042141}, which assumes that particles are self-propelled by assuming the magnitude of velocity to be a constant whereas its orientation is such that it aligns with that of surrounding particles. The Vicsek model is used for modeling the flocking of birds \cite{pnas0711437105,pnas1005766107,pnas1118633109}, schools of fish \cite{Tunstrm2013,Calovi_2014} and countless other collections of aligning active agents \cite{marchetti2013hydrodynamics,Gompper_2020,Palacci2013,Zhang2010,Sanchez2012}. To understand how a continuum description like \eqref{eq:tonertuequations} could ever arise, one can follow a kinetic theory approach. Several works have studied a Boltzmann equation that accounts for the interactions of microscopic models either identical or similar to the Vicsek model and coarse-grained it using a version of the Chapman-Enskog method \cite{chapman1990mathematical} that is suitable for self-propelling particles of constant velocity magnitude \cite{PhysRevE.74.022101,Bertin_2009,Ihle_2011,Ihle_2016,Lam_2015,PhysRevE.86.021120,Ihle_2011,PhysRevLett.109.098101,PhysRevX.4.041030,chen2023molecularchaosdenseactive,Mahault_2018,PhysRevE.111.044411}\footnote{In one recent work \cite{grosvenor2025hydrodynamicsboostinvariancekinetictheory}, the authors derived the Toner-Tu equation from the kinetic theory without starting from self-propelling particles but instead following an approach based on boost-agnostic hydrodynamics \cite{PhysRevE.110.054108,10.21468/SciPostPhys.9.2.018,10.21468/SciPostPhys.11.3.054,de_Boer_2018} and the relaxation time approximation \cite{PhysRev.94.511}. \textcolor{black}{What makes \cite{grosvenor2025hydrodynamicsboostinvariancekinetictheory} different from previous works based on kinetic theory for Vicseklike models as well as the present work, is that the $\alpha $-term does not arise from particle interactions but from a particular choice of the \quotes{equation of kinetic state}}.}
\newline 
For particles to maintain a constant velocity magnitude is incompatible with Galilean invariance and passivity. Therefore, such self-propelling particles already possess the essential features that allow the corresponding continuum description given by \eqref{eq:tonertuequations} to exhibit a flocking transition. Starting a coarse-graining procedure from this type of active microscopic dynamics is a natural choice, especially given the widespread use of Vicsek-like models. At the same time, the original raison d’être of kinetic theory was to show how a continuum description with an \textit{arrow of time} can emerge from simple microscopic interactions, which are typically reversible and symmetric \cite{Cercignani1988,degroot1984nonequilibrium,Cercignani2006-if}.
\newline 
These considerations inspire this work where we introduce a kinetic theory that is microscopically reversible \cite{onsager1931reciprocal,onsager1931reciprocala,RevModPhys.17.343,doi:10.1073/pnas.11.7.436,LIGHT1969281} and explicitly formulates the two essential ingredients that allow for the coefficient $\alpha$ in \eqref{eq:tonertuequations} to exist and have a negative sign. In particular, \textcolor{black}{the toy model} considered in this work does not disregard Galilean invariance or momentum conservation, but instead retains both properties by explicitly accounting for collisions with air particles that form the medium in which birds flap their wings. Secondly, activity is introduced by making the collisions of the birds with air reactive, allowing one to turn on a chemostat \cite{PhysRevE.60.2127,julicher2018hydrodynamic} that activates the system. As will become clear, deriving a continuum description comes with many challenges, the most important being that it is difficult to find a systematic expansion that allows one to efficiently extract all the coefficients in \eqref{eq:tonertuequations}. We therefore restrict to the computation of $\alpha $, which, as argued above, lies at the heart of the flocking transition. We consider a microscopic model based on two species of hard spheres which can undergo reactive collisions. Because of the symmetric nature of hard spheres as well as Galilean invariance, the nature of the collision is exclusively specified by the relative velocity and the impact parameter. We find that in order to bring about a sign flip of $\alpha$ which can give rise to a flocking transition, it is not enough to turn on activity and introduce medium from which to draw momentum. Instead, one also needs to fine-tune the nature of the bird-air collisions so that they are exclusively of the grazing kind, i.e. the impact parameter of the collision must be large. This is because frontal bird-air collisions strongly undermine the possibility of flocking.
\newline 
The structure of this work is as follows. In Sec.~\ref{sec:microscopicmodel} we describe our model formed by different types of hard sphere particles which can display binary collisions that can be reactive in a microscopically reversible way. We formulate the Boltzmann equations of the different particle species and derive the corresponding balance laws, show that $H$-theorem holds and derive the Maxwell-Boltzmann distributions and describe the state of chemical balance and its departure from equilibrium in the presence of a chemostat. In Sec.~\ref{sec:CG}, we formulate a coarse-grained description of momentum balance and extract the source term of the bird momentum balance equation whose leading order contribution will give the $\alpha$ coefficient of \eqref{eq:tonertuequations}. In Sec.~\ref{sec:flockingtransition}, we compute the numerical value of $\alpha$ as a function of temperature, affinity and the \quotes{grazing angle} and find that under specific circumstances, a sign flip can take place towards the negative sign so that this term becomes a growth rate, hinting at a flocking transition. \textcolor{black}{It is important to note that although the word \quotes{bird} is used very often in this work, we do not claim that the model considered in this work is a biologically faithful representation of avian flocking that occurs in Nature.}
\section{Microscopic model}
\label{sec:microscopicmodel}
\subsection{Bird collisions}
Let us now describe our microscopically reversible theory for bird flocking. The kinetic theory is one formed by a mixture of different types of particles which represent either birds or air. We consider a three-dimensional system composed of birds as well as air particles, both of which will be modeled as hard spheres. The birds can collide in three distinct ways, to wit
\begin{enumerate}
    \item Birds can collide with other birds
    \item Birds can flap their wings
    \item Birds can \quotes{unflap} their wings
\end{enumerate}
What we mean when we say that \quotes{birds can flap their wings} is that a bird collides with air in such a way that a chemical reaction occurs that transfers chemical energy to kinetic energy. Since we impose microscopic reversibility, we must also allow for the opposite collision, which one can call \quotes{unflapping}. Defining the components C$_{\pm}$ as the two types of chemicals, $\bird $ as bird and $\air$ as air, the reaction formula corresponding to the flapping or unflapping of wings is given by
\begin{align}  \label{eq:C1C2}
\text{\ce{ B +   A + C_+  <=> B  + A +  C_-   }} + \Delta E ~~ , 
\end{align}
where $\Delta E$ is the change in internal energy involved in the reversible reaction, \textcolor{black}{i.e. it is the energy per flap}. The reaction of \eqref{eq:C1C2} is taken to be exothermic in the forward direction, i.e. $\Delta E >  0 $. It follows that $\bird $ and $\air $ act merely as catalysts. The reason that such a specific reaction can represent the flapping of wings of a bird is because for a flap of wings to take place, energy must be consumed. But when this happens the wings also need to push off against the air, thereby allowing for the exchange of momentum between the environment and the bird. This fuel-consuming violation of bird momentum conservation is key to obtaining an $\alpha$-coefficient that can flip sign so that a flocking transition can take place. We now formulate a kinetic theory of reactive interspecies collisions that allows for the possibility of the reversible reaction of \eqref{eq:C1C2} to take place upon collision of a bird and an air particle. 
\subsection{Boltzmann equations}
To describe these three distinct bird collisions, let us consider the following Boltzmann equation for the bird phase space probability density $f_{\bird }$
\begin{align}  \label{eq:birdBoltzmmannequation}
   \frac{ \partial }{ \partial t  } f_{\bird}    +  \mathbf{v} \cdot  \frac{ \partial }{  \partial  \mathbf{x}  } f_{\bird}   =   \mathcal{C}^{\bird }_{\self }     +  \mathcal{C}^{\bird  }_{\flap }  ~~   , 
 \end{align}
 where $\mathcal{C}^{\bird }_{\self }$ represents bird-bird collisions and the $\mathcal{C}^{\bird  }_{\flap }$ represent flaps and unflaps with air. Similarly, there is a Boltzmann equation for the probability density of air particles $f_{\air }$ given by
 \begin{align}  \label{eq:airequation}
      \frac{ \partial }{ \partial t  } f_{\air}  +  \mathbf{w} \cdot  \frac{ \partial }{  \partial  \mathbf{x}  }  f_{\air}  =  \mathcal{C}^{\air }_{\self }      +\mathcal{C}^{\air}_{\flap }   ~~ .  
\end{align}
In \eqref{eq:birdBoltzmmannequation} and \eqref{eq:airequation} we omitted any force. It follows that the collision integrals are constrained by number density conservation as
\begin{subequations}
\begin{align}
  &     \int   d \mathbf{v}    \mathcal{C}^{\bird }_{\self }   =     \int   d \mathbf{w}    \mathcal{C}^{\air }_{\self }=0  ~~ , \\ 
   & \int   d \mathbf{v}    \mathcal{C}^{\bird    }_{\flap }     =  \int   d \mathbf{w} \mathcal{C}^{\air    }_{\flap }  =0  ~~  ,   \label{eq:equationflap}
   \end{align}
   as well as by the momentum balance as
   \begin{align}
   & m_\bird 
     \int   d \mathbf{v}  \, \mathbf{v}    \mathcal{C}^{\bird }_{\self }   =  
   m_\air     \int   d \mathbf{w} \, \mathbf{w}     \mathcal{C}^{\air }_{\self }    =0  ~~ ,  \\
    & m_\bird 
     \int   d \mathbf{v}  \, \mathbf{v}     \mathcal{C}^{\bird    }_{\flap }   +  
   m_\air     \int   d \mathbf{w} \, \mathbf{w}      \mathcal{C}^{\air  }_{\flap }   =0  ~~ ,
      \end{align}
\end{subequations}
where $m_{\bird , \air }$ are the bird and air particle mass respectively. Lastly, we have the equation for the spatial density $\rho^{\pm}$ of the chemicals $C_{\pm}$ given by
\begin{align}
  \frac{ \partial }{ \partial t  } \rho^{\pm}  =    \mathcal{C}^{\text{C} \pm      }_{\flap } ~~ .  
\end{align}
Note that we assume that the density $ \rho^{\pm}  $ does not have any spatial dynamics, and therefore $\rho^{\pm}$ is only a (uniform) density in real space. Number density conservation of the chemicals requires that
\begin{align} \label{eq:plusminusrelation}
    \sum_{\pm} \mathcal{C}^{\text{C} \pm      }_{\flap }   =0  ~~   , 
\end{align}
where we note that $\rho^{\pm}$ unlike $f_{\bird , \air }$ is only a density in real space but not in momentum space as we do not allow for spatial dynamics of the C$_\pm$ chemicals. The conservation of total energy thus dictates that
\begin{subequations}
    \begin{align}   
 &  \frac{1}{2} m_\bird 
     \int   d \mathbf{v}  \, v^2     \mathcal{C}^{\bird }_{\self }   =  
   \frac{1}{2}   m_\air     \int   d \mathbf{w} \, w^2      \mathcal{C}^{\air }_{\self }   =  0 ~~   ,  \\
  &   \frac{1}{2}  m_\bird 
     \int   d \mathbf{v}  \, v^2 \mathcal{C}^{\bird    }_{\flap }   +  
   \frac{1}{2}   m_\air     \int   d \mathbf{w} \, w^2     \mathcal{C}^{\air   }_{\flap }    =     \Delta E  \,   \mathcal{C}^{\text{C}  -       }_{\flap } ~~ .  \label{eq:energyconservation111}
\end{align}
\end{subequations}
\eqref{eq:energyconservation111} is proven in App.~\ref{app:energyproof} for the collision integrals introduced in the next section.

 \subsection{Bird-bird collisions}
\begin{figure}
    \centering
    \tikz[thick]{
        \node (img) at (0,0) {\includegraphics[width=0.8\linewidth]{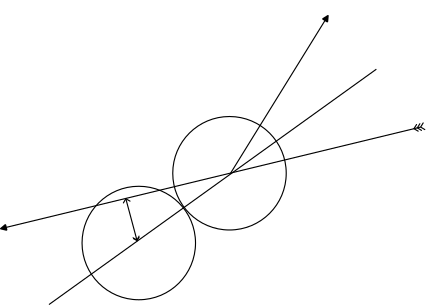}};
        
        \node at (2.5, -0.1) {$\mathbf{g}$};
        
        \node at (1.6, 1.4) {$\mathbf{g}'$};

        \node at (0.15, -0.05) {$\chi$};

        \node at (0.65, 0.2) {$\psi$};
        \node at (0.85, 0) {$\psi$};

        \node at (-1.6, -1.1) {$b$};
        
    }
    \caption{Depiction of a binary hard-sphere collision between two bird particles with incoming and outgoing relative velocities $\mathbf{g}$ and $\mathbf{g}'$, scattering angle $\chi$, and impact parameter $b$.}
    \label{figsimpleimpact}
\end{figure}

 Let us now discuss the precise nature of the two collision terms on the right-hand side of \eqref{eq:birdBoltzmmannequation}. The collision integral that describes how birds collide with themselves will be taken to be of the standard form for hard spheres undergoing binary collisions assuming molecular chaos \cite{Jeans_2009,grad1958principles,Cercignani1988}. That is, we have
 \begin{align}  \label{eq:selfcollision}
 \mathcal{C}^{\bird}_{\self }   =\frac{d^2_{\bird   }}{4} \iint d  \mathbf{v}_2  d \mathbf{e}'   \,   g      
  \left( f_{\bird} (\mathbf{v}^{\prime}_1 ) f_{\bird} (  \mathbf{v}^{\prime }_2   ) -      f_{\bird} ( \mathbf{v} ) f_{\bird} (  \mathbf{v}_2    )  )  \right)    &      ~~ .   \end{align}
Here we introduced the bird diameter $d_{\bird   }$, the bird velocity $\mathbf{v}$, and the relative velocity magnitude
\begin{align}
    g = | \mathbf{v} - \mathbf{v}_2 | ~~  . 
\end{align}
Lastly, we introduced the unit vector $\mathbf{e}'$ which represents the orientation of the relative velocity of the outgoing birds with respect to that of the ingoing birds. Its polar angle is the scattering angle $\chi$. The collision integral of \eqref{eq:selfcollision} is obtained by integrating over all the possible collisions with different impact parameters $b$ and relative velocity magnitude $g$ that can take place, and changing the integration variable of $b$ to $\chi$  by using the relation (see App.~\ref{eq:measurehardspheres} and Fig.~\ref{figsimpleimpact})
\begin{align}
    b = d_{\bird } \cos(\chi / 2  ) ~~ . 
\end{align}
The velocities $\mathbf{v}^{\prime }_2 $ and $\mathbf{v}^{\prime }_1 $ are tied to $\mathbf{v} $ and $\mathbf{v}_2 $ through momentum conservation and energy balance as
\begin{align}
    \mathbf{v} +     \mathbf{v}_2 =   \mathbf{v}_1' +     \mathbf{v}_2' ~~ , ~~  v^2 +     v^2_2 =   (v_1')^2  +     (v_2')^2  ~~ . 
\end{align}
 
 \subsection{Bird-air collisions}
 Let us now consider collisions between bird and air, which unlike the collision with birds among themselves involve a chemical reaction described by \eqref{eq:C1C2}. Like the birds, we assume the air particles are hard spheres. That is, we can integrate over all possible collisions specified by an impact parameter $b$ and a relative velocity magnitude 
\begin{align}
    u = | \mathbf{v} - \mathbf{w}_2 | ~~  . 
\end{align} 

\begin{figure}
\centering
\begin{tikzpicture}
  \node[anchor=south west,inner sep=0] (g) at (0,0)
    {\includegraphics[width=0.9\linewidth]{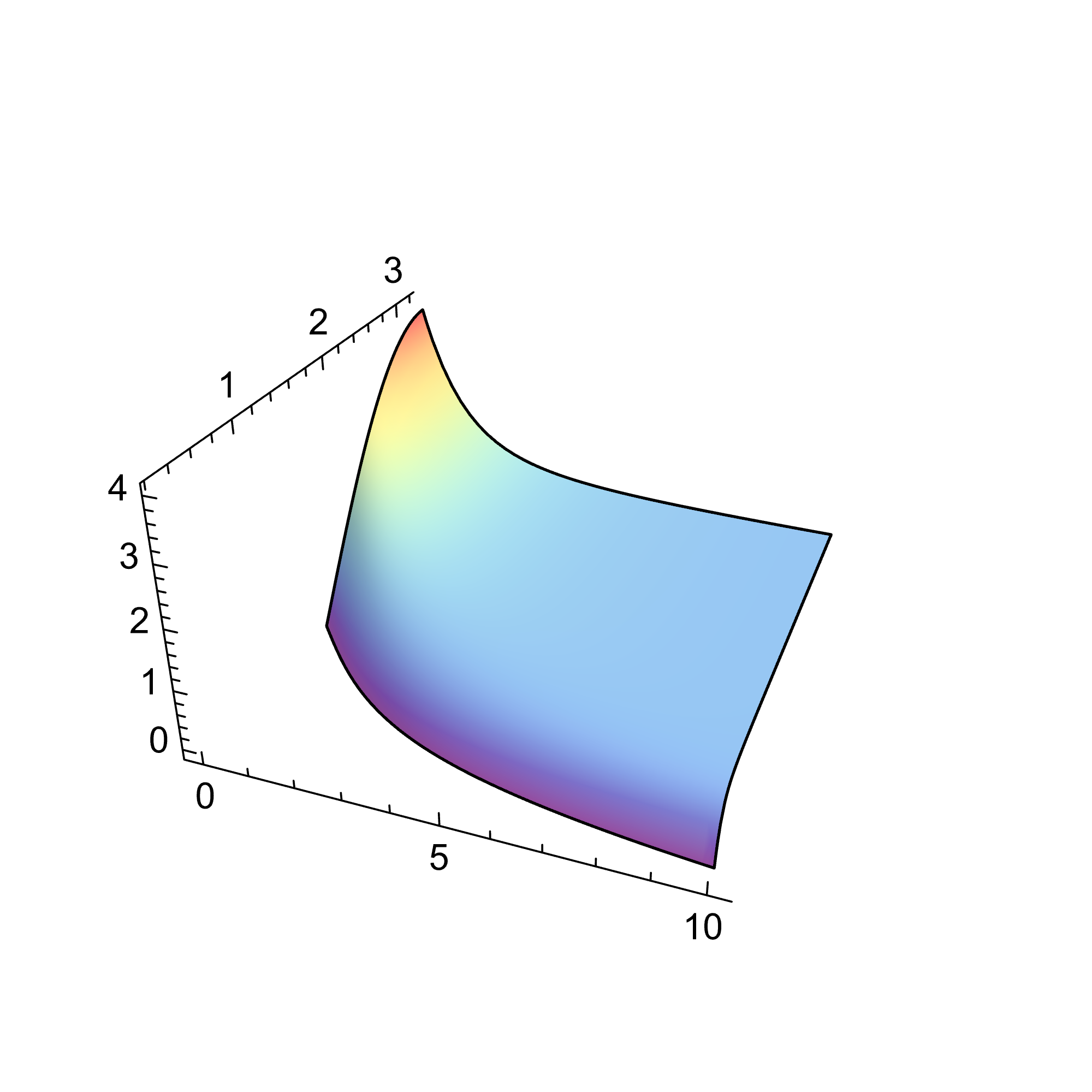}};
  \begin{scope}[x={(g.south east)},y={(g.north west)}, every node/.style={font=\Large}]
    \node[fill=white] at (0.4,0.15) {$u$};
    \node[fill=white] at (0.2,0.75) {$\chi$};
    \node[fill=white] at (-0.05,0.45) {$G^+(u,\chi)$};

  \end{scope}
\end{tikzpicture}

\caption{3D plot of $G^+(u,\chi)$ within the domain \eqref{eq:domain}, with $a=3$.}
\label{fig:enter-label123}
\end{figure}

\begin{figure*}[t]
\centering
\begin{tikzpicture}
  \node[anchor=south west,inner sep=0] (img) at (0,0)
    {\includegraphics[width=0.8\linewidth]{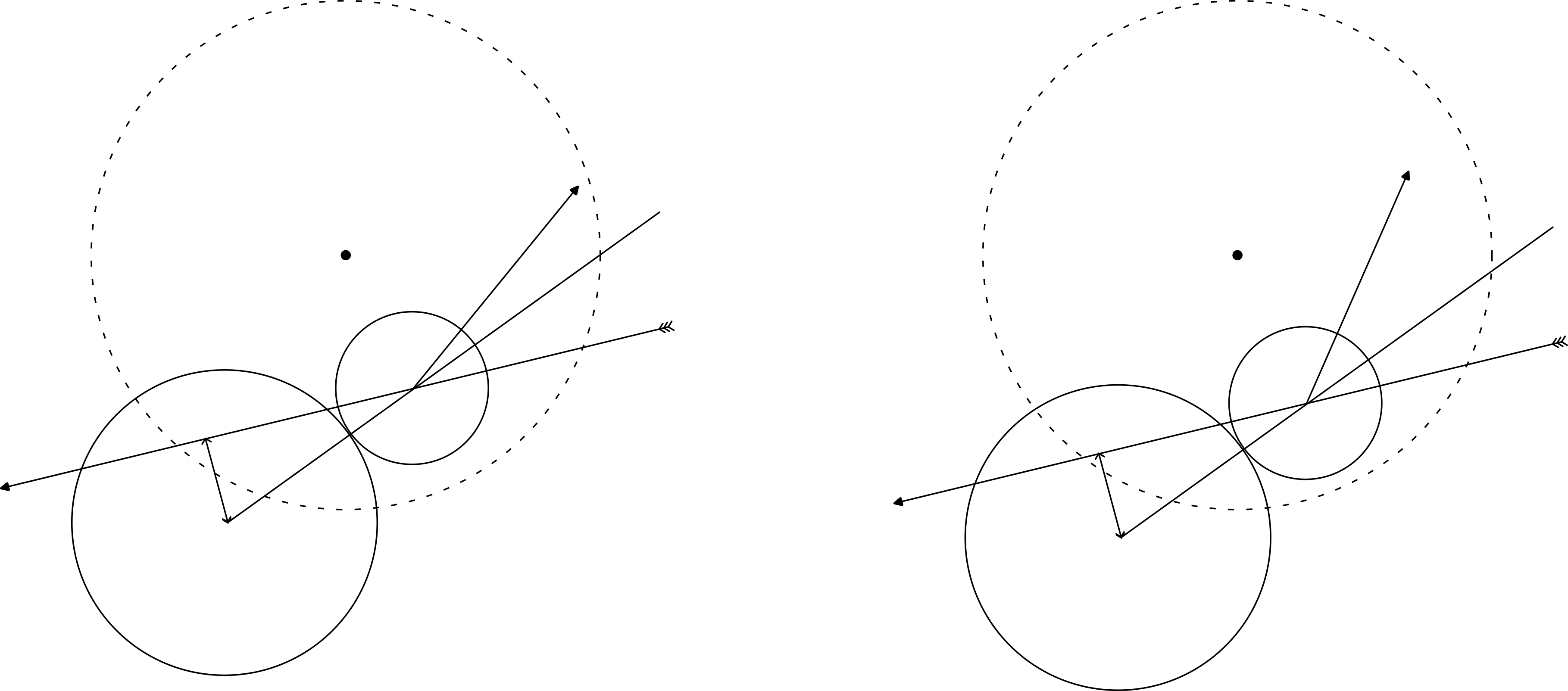}};
  \begin{scope}[x={(img.south east)},y={(img.north west)}]

    \node[fill=white, inner sep=2pt] at (0.08,0.92) {\bfseries (a)};
    \node[fill=white, inner sep=2pt] at (0.58,0.92) {\bfseries (b)};

    \node at (0.4,0.55) {$\mathbf g$};
    \node at (0.4-0.07,0.55+0.17) {$\mathbf g'$};

    \node at (0.35,0.52) {$\psi^{+}$};
    \node at (0.35,0.61) {$\psi^{\prime +}$};
 \node at (0.16,0.33) {$b^{+}$};

    \node[fill=white, inner sep=1.5pt] at (0.24,0.68) {$\kappa^{+}(u,\chi)$};
    \node[inner sep=1.5pt] at (0.26,0.5) {$\chi^{+}$};

  \node at (0.4+0.58,0.55) {$\mathbf g$};
    \node at (0.4-0.07+0.58,0.55+0.17) {$\mathbf g'$};
    \node at (0.16+0.58,0.33) {$b^{ -}$};
  \node at (0.35+0.58,0.52) {$\psi^{-}$};
    \node at (0.32+0.58,0.6) {$\psi^{\prime -}$};

    \node[fill=white, inner sep=1.5pt] at (0.24+0.58,0.68) {$\kappa^{-}(u,\chi)$};
    \node[inner sep=1.5pt] at (0.24+0.58,0.47) {$\chi^{-}$};

  \end{scope}
\end{tikzpicture}

\caption{Two-dimensional depiction of (a) a forward reactive bird-air collision and (b) an inverse reactive bird-air collision when the collision occurs inside the \quotes{reactive volume} $\kappa^{\pm}(u,\chi)$. The ingoing and outgoing relative velocities are drawn from the bird’s rest frame before and after the collision, respectively. For any forward or inverse reactive collision, $ \psi^{+}\ge \psi^{\prime +}$ and $\psi^{-}\le \psi^{\prime -}$.}
\label{figsimpleimpact123}
\end{figure*}

 Unlike for the case of bird-bird collisions, we assume that when a bird and air particle meet, the reactive collision only occurs with some probability less than one. \textcolor{black}{This probability can only depend on relative velocity magnitude $u$ and impact parameter $b$ by symmetry. That is, we have
 \begin{align}  \label{eq:condprob}
     P (\text{$\pm$ reaction} |  \text{encounter}  ,  u ,   b  ) =  \rho^{\pm} \tilde \kappa^{\pm}_{\flap } \left( u ,  b      \right) ~~ . 
 \end{align}
where we introduced the \quotes{reactive volumes}  $ \tilde \kappa^{\pm}_{\flap } \left( u ,   b    \right)$ which represent the spherical volume around a chemical $C_{\pm}$ for which the reaction of \eqref{eq:C1C2} occurs if a bird and air particle collide within it (see Fig.~\ref{figsimpleimpact123}). It will turn out to be more convenient to replace dependence on impact parameter with dependence on the scattering angle $\chi$ of the reactive collision which has a unique value for each impact parameter of a specific collision type. In particular, let us write
\begin{align}
    b = d_{\bird \air } g^{\pm}( u , \chi ) ~~  ,
\end{align}  
where $d_{\bird \air }$ is the average diameter of a bird and air particle and $g^{\pm}( u , \chi ) $ is a dimensionless function that is computed in App.~\ref{eq:scatteringangle}. We can then define 
\begin{align}
    \kappa^{\pm }_{\flap } \left( u  , \chi   \right)   =      \tilde \kappa^{\pm }_{\flap } \left( u  ,   d_{\bird \air } g^{\pm}( u , \chi )    \right)  ~~ .
\end{align}
The volumes $\kappa^{\pm}_{\flap } (u ,  \chi )$ are related to each other through microscopic reversibility as
(see App.~\ref{app:reactionprob} for a derivation)
    \begin{align}  \label{eq:alphaterms}
       \kappa^{-}_{\flap } \left( u  , \chi   \right)   =   \kappa^{+ }_{\flap } \left( \sqrt{u^2  -  a^2 }   , \chi   \right) \Theta ( u^2  -  a^2)  ~~ ,     
\end{align}}
where $\Theta ( x )  $ is the Heaviside step function and we introduced the shorthand notation
\begin{align}
  a^2    = \frac{2 \Delta E }{ \mur  } ~~ , ~~\mur  =  \frac{  m_\air m_{\bird }}{ m_\air + m_{\bird }}   ~~ , 
\end{align}
where $\mur$ is the \quotes{reduced mass} \cite{chapman1990mathematical} and $ m_\air $ and $m_{\bird }$ are the mass of the air and bird particle respectively. We assume that when bird and air particles do not react, they do not collide at all, meaning the bird manage to somehow \quotes{dodge} the air particles. That is, we assume
  \begin{align}  \label{eq:condprob12}
    \begin{split}    
     &  P (\text{dodge} |  \text{encounter}  ,  u ,   b    ) \\ 
     & =  1-   \sum_{\pm} P (\text{$\pm$ reaction} |  \text{encounter}  ,  u ,  b )   ~~ .
     \end{split}
 \end{align}
The reason we assume that bird particles are able to dodge air particles for certain values of $\chi$ and $u$ is that this turns out to be necessary to establish a flocking transition. Let us now consider the flapping collision integral, which can be split into a flapping and unflapping part so that we have
\begin{align}
    \mathcal{C}^{\bird }_{\flap } = \sum_{\pm} \mathcal{C}^{\bird, \pm }_{\flap } ~~ . 
\end{align}
We find these two parts to be given by (see App.~\ref{app:reactionprob} for a derivation)
\begin{widetext}
    \begin{align}  \label{eq:flapflapcoll}
\mathcal{C}^{\bird, \pm }_{\flap }    =  \frac{d^2_{\bird , \air }}{4} \iint  d  \mathbf{w}_2  d \mathbf{e}' \,       u \,   G^{\pm } ( u , \chi  )   \kappa^{\pm}_{\flap } \left( u   ,\chi   \right)    \left(   \rho^{\mp  }   
     f_{\bird} ( \mathbf{v}_1^{\mp \prime} ) f_{\air} (  \mathbf{w}^{\mp \prime}_2    ) -      \rho^{\pm  }     f_{\bird} (\mathbf{v}) f_{\air} (  \mathbf{w}_{2 }    )    \right)   &      ~~ , \end{align}
     \end{widetext}
Momentum conservation and energy balance require that
\begin{subequations}
\label{eq:energyconservation123}
    \begin{align}
   m_\bird  \mathbf{v} +   m_\air    \mathbf{w}_2  &  =  m_\bird   \mathbf{v}_1^{\prime \pm } +      m_\air  \mathbf{w}_2^{\prime \pm } ~~ ,  \\ 
  m_\bird  v^2 +   m_\air  v^2_2  & =    m_\bird   (v_1^{\prime \pm })^2  +     m_\air  (w_2^{\prime \pm })^2 \pm  2  \Delta E   ~~ . 
\end{align}
\end{subequations}
In App.~\ref{app:energyproof}, we show that the flapping collision integral of \eqref{eq:flapflapcoll} obeys the total energy conservation constraint given in \eqref{eq:energyconservation111}. The functions $G^{\pm  } ( u , \chi )$ in \eqref{eq:flapflapcoll} arise from substituting the impact parameter $b$ integration variable for the scattering angle $\chi$ of a reactive collision that does not conserve energy. The functions $G^{\pm  } ( u , \chi )$ are related as (see App.~\ref{eq:scatteringangle})
\begin{align}  \label{eq:Gthing}
G^- (u , \chi   )  = \Theta (u^2  - a^2 ) \frac{u^2  - a^2  }{ u^2  }   G^+ (\chi , \sqrt{u^2  - a^2 } )  ~~  . 
\end{align}
Their expressions are given by (see Fig.~\ref{fig:enter-label123} for a plot)
\begin{align}  \label{eq:Grelation}
    G  ^{ \pm } (u , \chi   )  & =  4 g^{\pm }  (u , \chi   )   \left( \left|\frac{\partial g^{ \pm  }  (u , \chi   )   }{ \partial \chi  }  \right| / \sin(\chi )  \right)   ~~. 
\end{align}
The function $g^{+} (u , \chi   ) $ is given by (see App.~\ref{eq:scatteringangle}) 
\begin{align} \label{eq:trickything2solve}
\begin{split}
    &  g^{ +   }  (u , \chi   )  =  \frac{\sin(\chi)}{\sqrt{ 1 +    \frac{u^2 }{  u^2   +   a^2     }  - 2  \frac{u}{   \sqrt{u^2   +   a^2  }   } \cos(\chi) }  }  .   
\end{split}
\end{align}
and $g^{ -  }  (u , \chi   )$ is related to $g^{ +   }  (u , \chi   )$ as
\begin{align}  \label{eq:simplification213444}
   g^- (u , \chi    )  = \Theta (u^2 - a^2) \frac{ \sqrt{u^2 -  a^2 } }{u}   g^+ ( \sqrt{u^2 - a^2 }  ,  \chi     ) ~~ ,  
\end{align}
as can be verified using \eqref{eq:Grelation}. The function $g^{ +   }  (u , \chi   )$ is only defined in the domain
\begin{align} \label{eq:domain}
    \arccos\left(  \frac{u}{   \sqrt{u^2   +   a^2  }   }    \right)    \leq  \chi \leq   \pi ~~ , 
\end{align}
whereas a similar finite $\chi$-domain for $g^{ -   }  (u , \chi   )$ can be inferred from \eqref{eq:simplification213444}. \eqref{eq:domain} means that, unlike is the case for nonreactive collisions, not every scattering angle is possible for a reactive collision. This is because, in an inverse reactive collision, the scattering angle must not be too small, as the particles would only graze each other to the point where they are not be able to provide the desired amount of ingoing kinetic energy needed for the reaction to take place. Conversely, the outgoing kinetic energy of the forward reaction moves the scattering angle away from zero when the ingoing particles only graze each other.

\subsection{$H$-theorem}
Before we make the system active by turning on the chemostat, let us see how the distribution functions move towards equilibrium in the absence of activity as is required by the Second Law of Thermodynamics. To understand the Second Law in the context of the Boltzmann equation, we consider the $H$-functional which is composed of the bird and air contribution
\begin{equation}
H_{\inv}  = \iint d \mathbf{r} d  \mathbf{v} f_{\inv} \log f_{\inv}    ~~ , ~~  \inv \in \{ \air  , \bird \}   , 
\end{equation}
whereas the chemical contribution to the $H$-functional is given by 
\begin{equation}
H^{\pm}_{\rho } =  \int d \mathbf{r}   \rho^{\pm} \log \rho^{\pm}   , 
\end{equation}
The time evolution of $H$ is given by
\begin{align}
\begin{split}
\frac{\partial }{ \partial t} H  =  &  \sum_{\inv   }  \iint d \mathbf{r}  d\mathbf{v}  \frac{\partial f_{\inv}}{\partial t} \left( 1 + \log f_{\inv} \right) \, \\  & +   \sum_{\pm}  \int d \mathbf{r} \frac{\partial    \rho^{\pm}  }{\partial t }  (1 + \log (\rho^{\pm }   )   )    .
\end{split}
\end{align}
$H$-theorem dictates that in equilibrium it holds that
\begin{align} \label{eq:Htheorem}
    \frac{\partial }{ \partial t} H^{(0)}  =  \frac{\partial }{ \partial t} H^{(0)}_{\self } +  \frac{\partial }{ \partial t} H^{(0)}_{\flap  } =0  ~~ ,
\end{align}
where $H^{(0)}$ is the $H$-functional in equilibrium and we have disregarded the possibility of a state inhomogeneous in space for simplicity. Using \eqref{eq:birdBoltzmmannequation} we find
\begin{equation}  \label{eq:Hself}
\frac{\partial}{ \partial t}  H_{\self  } = \sum_{\inv}  \iint d \mathbf{r} d\mathbf{v}  \mathcal{C}^{I     }_{\self }   \log f_{\inv} ~~    .
\end{equation}
and
\begin{align}  \label{eq:flapequation}
\begin{split}
 \frac{\partial}{ \partial t}  H_{\flap  } &= \sum_{\pm , \inv }    \iint d \mathbf{r} d\mathbf{v}     \mathcal{C}^{I   \pm   }_{\flap }  \log f_{\inv}   \\ 
 & +  \sum_{\pm}     \int d \mathbf{r}   \mathcal{C}^{\rho     \pm   }_{\flap }   \log (\rho^{\pm }  )    .
\end{split}
\end{align}
It turns out to be possible to find an equilibrium state where the \eqref{eq:Hself} and \eqref{eq:flapequation} vanish independently, that is
\begin{align}  \label{eq:derivatives}
   \frac{\partial }{ \partial t} H^{(0)}_{\self } =  \frac{\partial }{ \partial t} H^{(0)}_{\flap  } =0  ~~ , 
\end{align}
 To prove \eqref{eq:derivatives}, let us introduce the Maxwell-Boltzmann distributions given by
\begin{subequations} \label{eq:maxwelldistirbutions}
    \begin{align}  
    f^{(0) }_{\bird} (\mathbf{v})   & = n_{\bird } \left(  \frac{m_{\bird }}{2 \pi \kb  T_{\bird } } \right)^{\frac{3}{2}}  \exp( -\frac{m_{\bird }}{2 \kb T_{\bird}  }  V^2    )  \\ 
    f^{(0)}_{\air} ( \mathbf{w})  & =  n_{\air } \left(  \frac{m_{\air }}{2 \pi \kb  T_{\air  } } \right)^{\frac{3}{2}}  \exp( -\frac{m_{\air }}{2 \kb T_{\air}  }  W^2     ) ~~ , 
\end{align}
\end{subequations}
where $\mathbf{V} = \mathbf{v} - \mathbf{v}_0 $ and $\mathbf{W} = \mathbf{w} - \mathbf{w}_0 $. Upon plugging in \eqref{eq:maxwelldistirbutions}, the vanishing of $\frac{\partial }{ \partial t} H^{(0)}_{\self } $ follows from the collision being conservative in momentum and energy \cite{chapman1990mathematical}. Imposing $\frac{\partial }{ \partial t} H^{(0)}_{\flap } =0 $ with \eqref{eq:maxwelldistirbutions} plugged in, we find the constraint
\begin{align} \label{eq:VsqWsq}
   &   \frac{m_{\bird }}{2 \kb T_{\bird}  }  V^2  + \frac{m_{\air }}{2 \kb T_{\air}  }  W^2  \\  &  = \frac{m_{\bird }}{2 \kb T_{\bird}  }   ( V^{\prime \pm } )^2   + \frac{m_{\air }}{2 \kb T_{\air}  }   ( W^{\prime \pm } )^2- \log(\rho_{\pm} / \rho_{\mp }) ~~ ,  
\end{align}
If we impose that $ \mathbf{v}_0 =  \mathbf{w}_0$ and $T_{\air} = T_{\bird}$, it follows from \eqref{eq:energyconservation123} that we have
\begin{align} \label{eq:VsqWsq123}
     \begin{split}
              &   m_{\bird }  V^2  +  m_{\air }  W^2   =  m_{\bird }   ( V^{\prime \pm } )^2   + m_{\air }   ( W^{\prime \pm } )^2  \mp  2  \Delta E   ,  \end{split}  \end{align}
so that \eqref{eq:VsqWsq} is obeyed when
\begin{align}   \label{eq:chemicalequilibrium}
  \rho^{+} =  \rho^{ -  }  \exp( - \beta_\air  \Delta E    ) ~~ , 
\end{align}
where $\beta_{\air }  = (\kb T)^{-1}$. Having specified the equilibrium state, let us prove that
\begin{align}
     \frac{\partial   }{ \partial t} H \leq  0 ~~ .   \label{eq:Htheoremsecond}
\end{align}
In App.~\ref{app:Htheorem}, we show that \eqref{eq:flapequation} is of the form  
\begin{align}
     \frac{\partial }{ \partial t} H_{\flap}  \sim    \log(x/y) ( y -x  ) \leq 0   ~~ , ~~ x , y  \geq 0  
\end{align}
Since it is well known that \eqref{eq:Hself} can be written in a similar way \cite{chapman1990mathematical,Cercignani1988,tong2012kinetic}, we find that \eqref{eq:Htheoremsecond} holds and it thus follows that in the absence of activity, the system will always be pushed towards the equilibrium state specified with  $\mathbf{v}_0 =  \mathbf{w}_0$ and $T_{\air} = T_{\bird}$ and \eqref{eq:chemicalequilibrium}.

\section{Coarse-graining the microscopic model}
\label{sec:CG}
\subsection{Bird balance equations}
We now discuss the balance equations for the bird density and energy that follow from taking moments of \eqref{eq:birdBoltzmmannequation}. 
Firstly we have conservation of number of birds:
\begin{align}  \label{eq:particleconservation}
     \frac{\partial}{\partial t } n_{\bird } +  
    \mathbf{\nabla} \cdot  (n_{\bird }   \mathbf{v}_0  ) &   =0 ~~ ,  
\end{align}
where $n_{\bird}   =     \int d \mathbf{v} f_{\bird} $ is the density of birds. Using \eqref{eq:particleconservation}, it follows that momentum balance is given by
\begin{align}
n_\bird m_{\bird}  
 \left( \frac{\partial}{\partial t}     \mathbf{v}_0       +   \mathbf{v}_0 \cdot  \nabla    \mathbf{v}_0  \right)     +\nabla \cdot  \mathbf{P}  = n_\bird m_{\bird} \mathbf{S}  ~~ ,     \label{eq:momentum}
\end{align}
where the average bird velocity $\mathbf{v}_0$ is given by
\begin{align}
 \mathbf{v}_0 =  \langle  \mathbf{v} \rangle       ~~ , 
\end{align}
and we introduced the source term $\mathbf{S}$ and stress tensor $\mathbf P $
\begin{subequations} \label{eq:plugingterms}
    \begin{align}  \label{eq:sourceterm}
   \mathbf{S}   & =  \frac{1}{n_{\bird } }  \int d \mathbf{v} \, \mathbf{v}  \mathcal{C}^{\bird}_{\flap  }   ~~ ,  \\     \mathbf{P}   &   = n_{\bird} m_{\bird } \left( \langle    \mathbf{v}   \mathbf{v}   \rangle  -   \mathbf{v}_0    \mathbf{v}_0 \right) ~~  , 
\end{align}
\end{subequations}
as well as the averaging notation
\begin{align}
    \langle ... \rangle = \frac{1}{n_{\bird }}  \int d \mathbf{v} f_{\bird} \left(  ...  \right)  ~~ . 
\end{align}
The term on the right-hand side of \eqref{eq:momentum} represents the violation of bird momentum conservation due to collisions with air. From the Boltzmann equation for air given by \eqref{eq:airequation}, one can derive a similar source term that would cancel the source term in \eqref{eq:momentum} so that the conservation of total momentum is confirmed. 

\subsection{Away from equilibrium}
\textcolor{black}{Now that we understand the equilibrium state to which the two-fluid system relaxes in the absence of thermodynamic forces, we consider the possibility that the variables that describe the equilibrium state of \eqref{eq:maxwelldistirbutions} are such that the system is not in global equilibrium. This is typically done through a Chapman-Enskog expansion \cite{chapman1990mathematical} where a state like that of \eqref{eq:maxwelldistirbutions} is assumed to be in local equilibrium, thus allowing for nonzero gradients of the hydrodynamic variables $T_\air$ and the mass conserving combination of $\mathbf{v}_0$ and $\mathbf{w}_0$.} However, the variable that must be turned on in order to find the coefficient $\alpha$ in \eqref{eq:tonertuequations} is the average relative velocity
\begin{align}
\mathbf{u}_0 =     \mathbf{v}_0  - \mathbf{w}_0       ~~ .   
\end{align}
As $\mathbf{u}_0$ is not associated with any conservation law, it is not a hydrodynamic variable. For passive matter, the variable $\mathbf{u}_0$ should therefore be discarded when considering the hydrodynamic regime as it is short-lived \cite{glorioso2018lectures}. However, in the presence of activity, this variable can get \quotes{activated}. This happens when the relaxation rate of $\mathbf{u}_0$ flips sign and becomes a growth rate that drives the flocking transition. As will be shown, the coefficient that can undergo this sign flip and thereby drive the flocking transition is the coefficient $\alpha$ in \eqref{eq:tonertuequations}. \textcolor{black}{Under what conditions $\alpha$ attains a negative sign within our model is precisely the question that we try to answer. \newline 
In order for the variable $\mathbf{u}_0$ to be activated, a fuel source is required. This role is played by the chemicals $C_{\pm}$ which in biochemical systems correspond to ATP and ADP respectively. We neglect any dynamics of the chemicals $C_{\pm}$ and instead assume that there is a chemical reservoir that maintains a steady chemical imbalance. This reservoir thus acts as a \quotes{chemostat}. The chemostat is quantified by a chemical potential difference $\Delta \mu$, which is also called reaction affinity \cite{degroot1984nonequilibrium,julicher2018hydrodynamic,PhysRevE.60.2127}. When $\Delta \mu > 0$ the chemostat induces a steady inflow of $C_+$ and outflow of $C_-$, whereas the opposite is true when $\Delta \mu < 0$. In this steady state of chemical imbalance, the chemical densities are related as }
\begin{align} \label{eq:chemicalequilibrium123}
    \rho^{+ } = \rho^{ - } \exp(   \beta_\air  (  \Delta \mu     - \Delta E  ) )  ~~ . 
\end{align}
\textcolor{black}{As long as $\Delta \mu \neq 0$, the system is out of equilibrium and it is therefore not strictly valid to assume that the particles are distributed according to \eqref{eq:maxwelldistirbutions} at leading order. However, we assume that for $\Delta \mu$ sufficiently close to zero, \eqref{eq:maxwelldistirbutions} can reasonably describe the distribution of bird and air particles \cite{julicher2018hydrodynamic}. Therefore, we compute the coefficient $\alpha$ by starting from \eqref{eq:maxwelldistirbutions}, taking
\begin{align}
T_{\bird} = T_{\air} ~~  . \label{eq:equaltemperature}    
\end{align}
It would also be possible to take $T_{\air} \neq T_{\bird}$, which would allow one to study the sourced bird energy balance equation, which is affected both indirectly by activity but also directly through the inflow of chemical energy \cite{armas2025hydrodynamicsthermalactivematter}. Since dynamics of temperature is not something that is described by the Toner-Tu equations of \eqref{eq:tonertuequations}, it is beyond the scope of this work.}

\subsection{$\alpha $-coefficient}
Let us now compute the coefficient $\alpha $. Plugging \eqref{eq:chemicalequilibrium123} and \eqref{eq:maxwelldistirbutions} with \eqref{eq:equaltemperature} into the flapping collision integral of \eqref{eq:flapflapcoll} and using \eqref{eq:energyconservation123}, we find
\begin{widetext}
    \begin{align}  
\begin{split}
\label{eq:frictionaxxxxx6123}  \mathcal{C}^{(0),\bird}_{\flap }    &  = \frac{d^2_{\bird , \air }}{4} \sum_{\pm} \rho^{\mp}      \iint d  \mathbf{w}_2  d \mep   \,   u         f^{(0)}_{\bird} ( \mathbf{v} ) f^{(0)}_{\air} ( \mathbf{w}_2)   G^{\mp} (u , \chi   ) \kappa^{\mp}_{\flap} (u  , \chi  )   \left(  \exp\left( 
 \beta_{\air }   
 \left(    \mur ( \mathbf{u}^{ \pm  \prime }  -  \mathbf{u} )   \cdot \mathbf{u}_0   \pm   \Delta \mu   \right) 
 \right)  -  1  \right)         ~~  . 
 \end{split}
 \end{align}
\end{widetext}
Having specified the collision integral, we can use it to compute the source term of momentum balance given by \eqref{eq:sourceterm}. \textcolor{black}{Let us expand this source term up to linear order $\mathbf{u}_0$, i.e.  \begin{align}  \label{eq:alphacoefficient}
    \mathbf{S} = - \alpha^{\prime } \mathbf{u}_0 + \mathbf{O} (\mathbf{u}_0^3) ~~ , 
\end{align}
where $\alpha^{\prime}$ is given by (see App.~\ref{app:sourceterm} for a derivation)}
 \begin{widetext}
       \begin{align}
     \begin{split} \label{eq:flapping}
               \alpha^{\prime }    =  &  -  \frac{1  }{3 }   \beta_\air  \mur    K \sum_{\pm} \rho^{\mp}   \iint    d u    d \chi \sin(\chi )       \,   u^5     G^{\mp} (u , \chi   ) \kappa^{\mp}_{\flap} (u  , \chi  )     \left(   \frac{ \sqrt{u^{2} \mp  a^2  }   }{u }     \cos(\chi ) \exp(    \pm \beta_{\air}  \Delta \mu       )  - 1  \right)  \exp( -\frac{   \mur   u^{ 2 }   }{2 \kb T_{\air}  }     )  \\  
   &  +       K  \sum_{\pm}  \rho^{\mp}    \iint  d  u   d \chi \sin(\chi ) \,   u^3        G^{\mp} (u , \chi   ) \kappa^{\mp}_{\flap} (u  , \chi  )       \left(  \exp\left( 
    \pm \beta_{\air}  \Delta \mu \right)  -  1  \right)      \exp( -\frac{ \mur   u^{ 2 }   }{2 \kb T_{\air}  }     )   ~~   ,        \end{split}
 \end{align}
\end{widetext}
where we introduced the constant
  \begin{align}
     K = \frac{   ( 2  \pi)^2  d^2_{\bird , \air }  n_{\air } }{2}  \frac{  m_\air   }{m_0  }    \left(  \frac{  \mur  \beta_\air }{ 2 \pi } \right)^{\frac{3}{2}}  ~~. 
 \end{align}
For simplicity, we now assume that the mass density of air is much higher than that of the birds so that we are justified in discarding the dynamics of air. We then  choose to work in the frame where it holds that $\mathbf{w}_0 =0 $. It then follows from \eqref{eq:momentum} that $\alpha^{\prime }$ is identical to that of the Toner-Tu equations given in \eqref{eq:tonertuequations}, i.e. it holds that
\begin{align}
    \alpha^{\prime } = \alpha ~~ . 
\end{align}
Now that we understand how the $\alpha$-term in \eqref{eq:tonertuequations} can follow from \eqref{eq:momentum}, it is natural to ask whether other terms in \eqref{eq:tonertuequations} could be derived in a similar way. In principle, any term in \eqref{eq:tonertuequations} could arise from \eqref{eq:momentum}, where it must be noted that the terms in \eqref{eq:tonertuequations} that are associated with the violation of Galilean invariance must come from the source term on the right hand side of \eqref{eq:momentum}. To obtain these terms, in particular the gradient terms, requires that one follows an approach similar to the Chapman-Enskog expansion \cite{chapman1990mathematical} where one iteratively solves the Boltzmann equation in such a way that the Boltzmann distribution gets corrections whose effect can be understood by plugging them back into \eqref{eq:plugingterms}. In this work, we restrict ourselves to the coefficient $\alpha$ which lies at the heart of the flocking transition.

\section{Growth rate}
\label{sec:flockingtransition}
\textcolor{black}{Having fully specified the microscopically reversible kinetic theory of flocking, let us now compute the coefficient $\alpha$, which, as argued in the introduction, is key to bringing about the flocking transition and furthermore captures the active and Galilean symmetry breaking characteristics of the Toner-Tu equations. We will do this in two steps. First, we do it for the case where the energy per flap $\Delta E =0$, because when this is so the system dramatically simplifies and becomes effectively passive, thus allowing one to precisely see why a continuum of passive birds could never display flocking. We furthermore s how frontal collisions, which are collisions with a scattering angle $\chi$ close to $\pi$, contribute more to the positivity of $\alpha$ than grazing collisions. With these lessons, we then compute the growth rate for energy per flap $\Delta E \neq 0$, and find that provided we filter out frontal collisions, turn on the chemostat and sufficiently cool down the system, $\alpha$ can flip sign and thus a flocking transition is possible.}
\subsection{Case $\Delta E =0 $}
When $\Delta E =0 $, it follows from \eqref{eq:alphaterms} that
\begin{align}
    \kappa_{\flap  , \Delta E  =0}^{+} (u , \chi ) =  \kappa^{-}_{\flap  , \Delta E  =0} (u, \chi ) \equiv   \kappa_{\flap , \Delta E  =0 } (u, \chi )  ~~ . 
\end{align}
Furthermore, it holds that $G_{\Delta E  =0}^{\pm } ( u , \chi ) =1 $ and \eqref{eq:chemicalequilibrium123} reduces to
\begin{align} \label{eq:chemicalequilibrium124}
    \rho^{+ }  \exp( -   \beta_\air  \Delta \mu     )  =    \rho^{- }   ~~ . 
\end{align}
The formula for $\alpha $ of \eqref{eq:flapping} simplifies to     \begin{widetext}
       \begin{align}
     \begin{split} \label{eq:flappingterm}
                \alpha_{\Delta E  =0  }    =  &    \frac{1  }{3 }   \beta_\air  \mur    K \sum_{\pm} \rho^{\mp}   \iint    d u    d \chi \sin(\chi )       \,   u^5      \kappa_{\flap , \Delta E  =0 } (u  , \chi  )     \left( 1 -  \cos(\chi )  \right)  \exp( -\frac{   \mur   u^{ 2 }   }{2 \kb T_{\air}  }     )    ~~   ,        \end{split}
 \end{align}
\end{widetext}
$\alpha_{\Delta E  =0  }$ is always positive, which tells us that the chemostat cannot induce a flocking transition when the chemical reaction to which the chemostat corresponds does not involve an energy transfer. This is expected, because in the absence of this energy transfer it is impossible to tell that a chemical reaction has taken place at all, rendering the system effectively passive. When the system is passive, $H$-theorem tells us that the coefficient $\alpha$ should always be such that $\mathbf{u}_0$ gets pushed back to zero. \fl{We additionally learn from \eqref{eq:flappingterm} that the integrand is most positive and thus most stabilizing when the value of $\chi $ is close to $\pi$, which corresponds to the statement that frontal collisions contribute more to stabilizing the bird-air mixture than grazing collisions do. \textcolor{black}{The reason for this is that a frontal collision changes velocity in a way that completely goes against the average velocity of birds with respect to air when the birds are in a flocking state.} This insight will be important when trying to bring about a flocking transition for the case $\Delta E > 0 $. }

\subsection{Case $\Delta E  >  0 $}
We now consider the evaluation of $\alpha$ for $\Delta E  >  0 $. In order to do this, we need to fully specify the nature of the reactive collisions, i.e. to specify $\rho^{\pm}$ and $\kappa^{\pm} ( u , \chi ) $. Firstly, since the sum of $\rho^{\pm}$ is conserved, we have
\begin{align}
    \rho^{\pm} = \frac{\rho^{\text{tot}}  }{ 1  + \exp( \pm   \beta_{ \air } ( \Delta E -  \Delta \mu   ) )    } ~~   . 
\end{align}
\textcolor{black}{When one wishes to have a realistic description of reactive particles, it is best to let $\kappa^{\pm } (u  , \chi )$ be monotonically increasing in $u$ beyond some threshold that corresponds to the activation energy \cite{Presentpaper,PRIGOGINE1949913,10.1063/1.1725565,10.1063/1.1731889,kugerl,LIGHT1969281}.} However, we are describing a toy model that aims to show how a flocking transition could arise from a microscopically reversible theory of reactive particles in the presence of a chemostat. Therefore, we will turn on reactions even for the smallest values of $u$ that are energetically allowed by taking 
\begin{align} \label{eq:rhorho}
\begin{split}
       \rho^{\text{tot}} \kappa^{+} (u , \chi   )  &  =    F ( u , \chi )   ~~ ,  \\  \rho^{\text{tot}} \kappa^{-} ( u , \chi   )   & =     \Theta (u^2  - a^2  ) F (\sqrt{u^2 - a^2 }  ,\chi )   ~~ , 
\end{split}
\end{align}
where the function $F(u ,  \chi )  $ is used to specify the $\chi $-domain. As follows from \eqref{eq:domain}, the domain of $\chi$ has a lower bound which reflect the fact that certain scattering angles are not consistent with a reactive collision. To fully specify $F(u , \chi )$ we make the simple choice
\begin{align}  \label{eq:choice}
F ( u  ,   \chi   )  =  \Theta \left(  \arccos\left(     \frac{u}{   \sqrt{u^2  +   a^2  }   }    \right)  + \chi_{\graze } - \chi  \right)      ~~ , 
\end{align}
so that the $\chi$-domain is also upper bounded and given by
\begin{align}
     0 \leq  \chi -  \arccos\left(  \frac{u}{\sqrt{u^2  +   a^2  } }    \right) \leq \chi_{\graze } ~~ . 
\end{align}

\begin{figure}[H]
\centering
\begin{tikzpicture}
  \node[anchor=south west,inner sep=0] (img) at (0,0)
    {\includegraphics[width=0.6\linewidth]{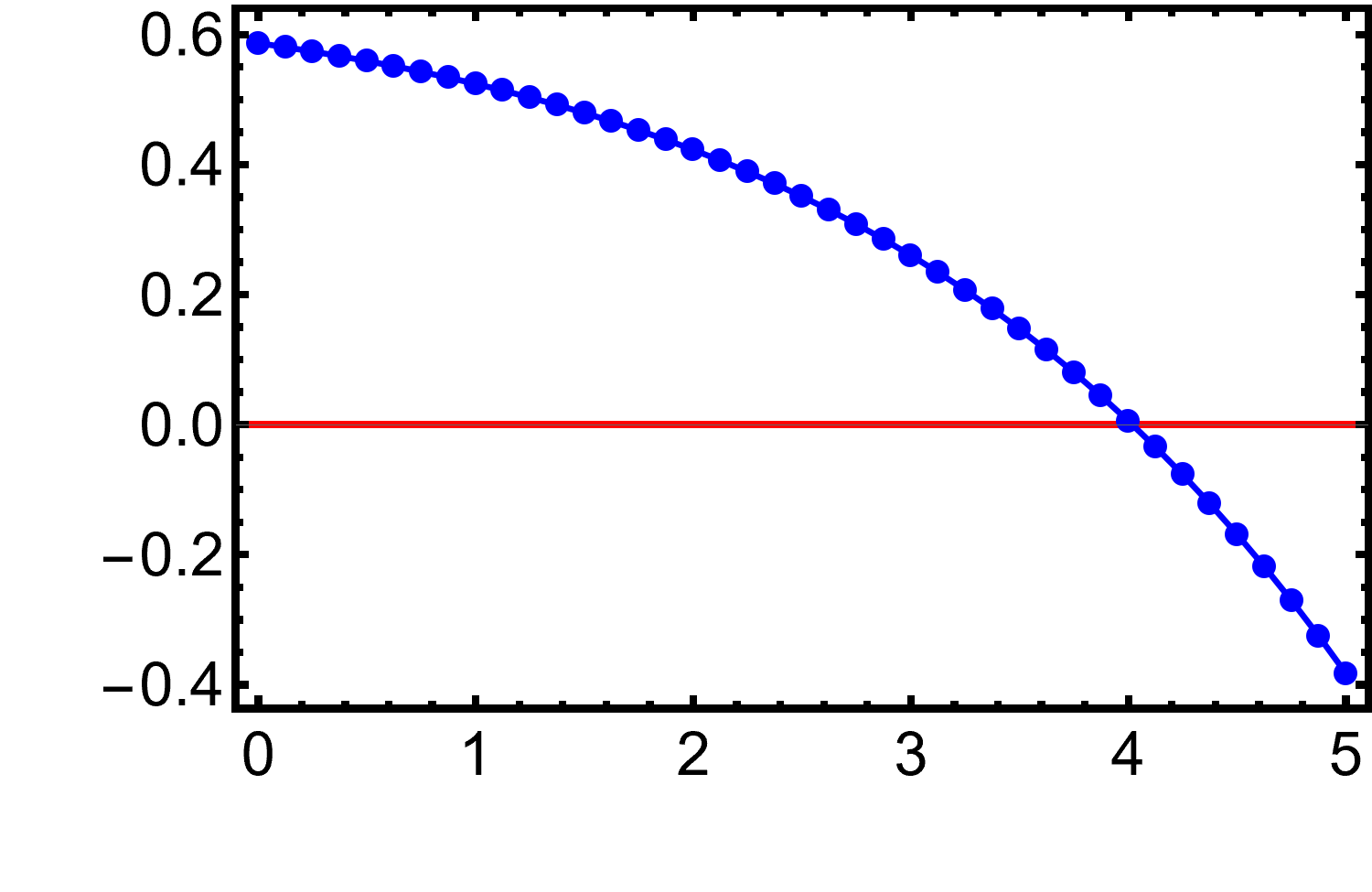}};
  \begin{scope}[x={(img.south east)},y={(img.north west)}]

    \node[fill=white] at (0,0.95) {$\text{(a)}$};
    \node[fill=white] at (0.05,0.65) {$\alpha$};
      \node[fill=white] at (0.6,0)  {$ \Delta \mu  $};
  \end{scope}
\end{tikzpicture}
\hfill
\begin{tikzpicture}
  \node[anchor=south west,inner sep=0] (img) at (0,0)
    {\includegraphics[width=0.6\linewidth]{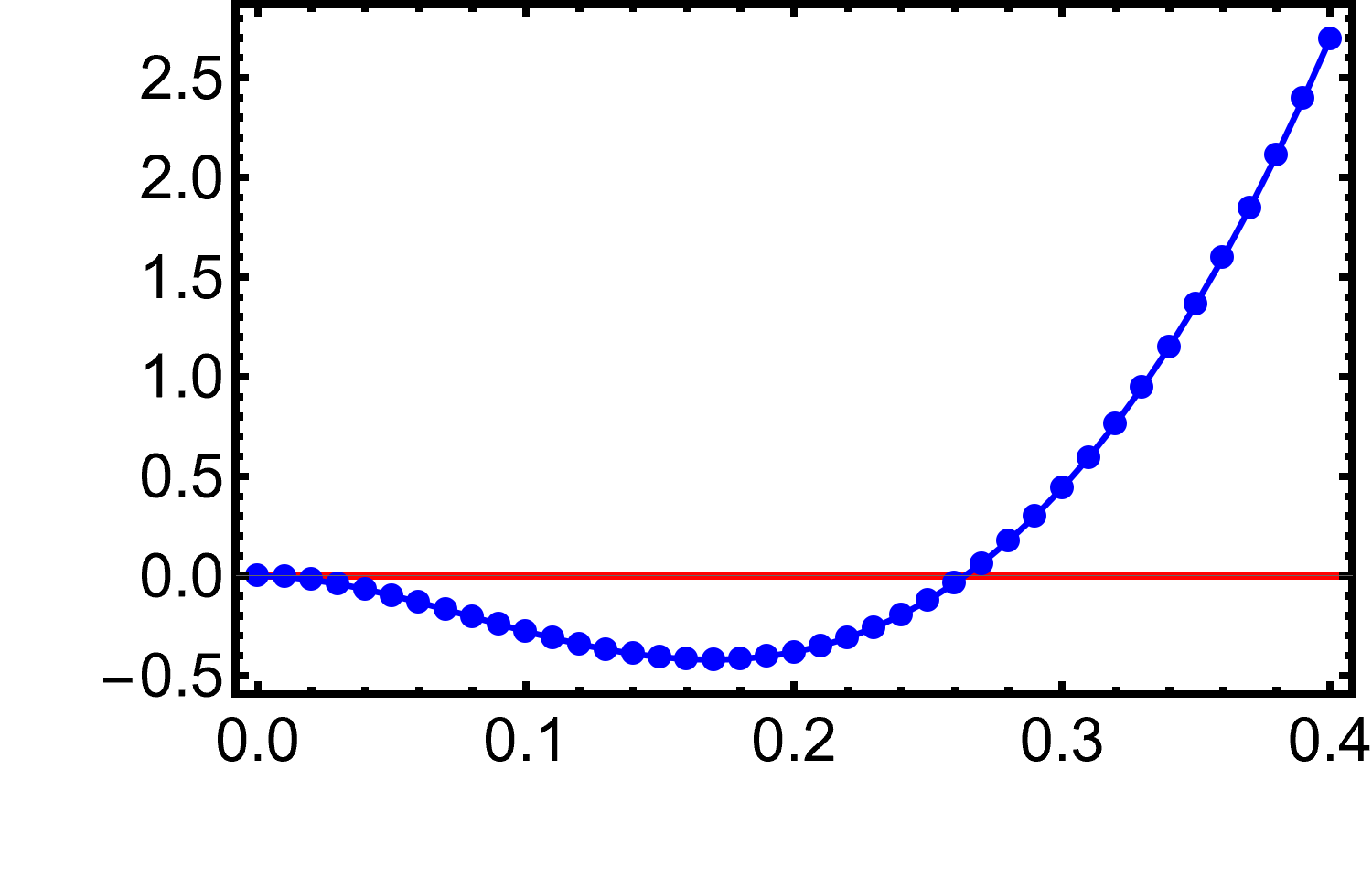}};
  \begin{scope}[x={(img.south east)},y={(img.north west)}]

    \node[fill=white] at (0,0.95) {$\text{(b)}$};
  \node[fill=white] at (0.6,0)  {$ \chi_{\graze} $};
    \node[fill=white] at (0.05,0.65) {$\alpha$};

  \end{scope}
\end{tikzpicture}
\vspace{1em}
\begin{tikzpicture}
  \node[anchor=south west,inner sep=0] (img) at (0,0)
    {\includegraphics[width=0.6\linewidth]{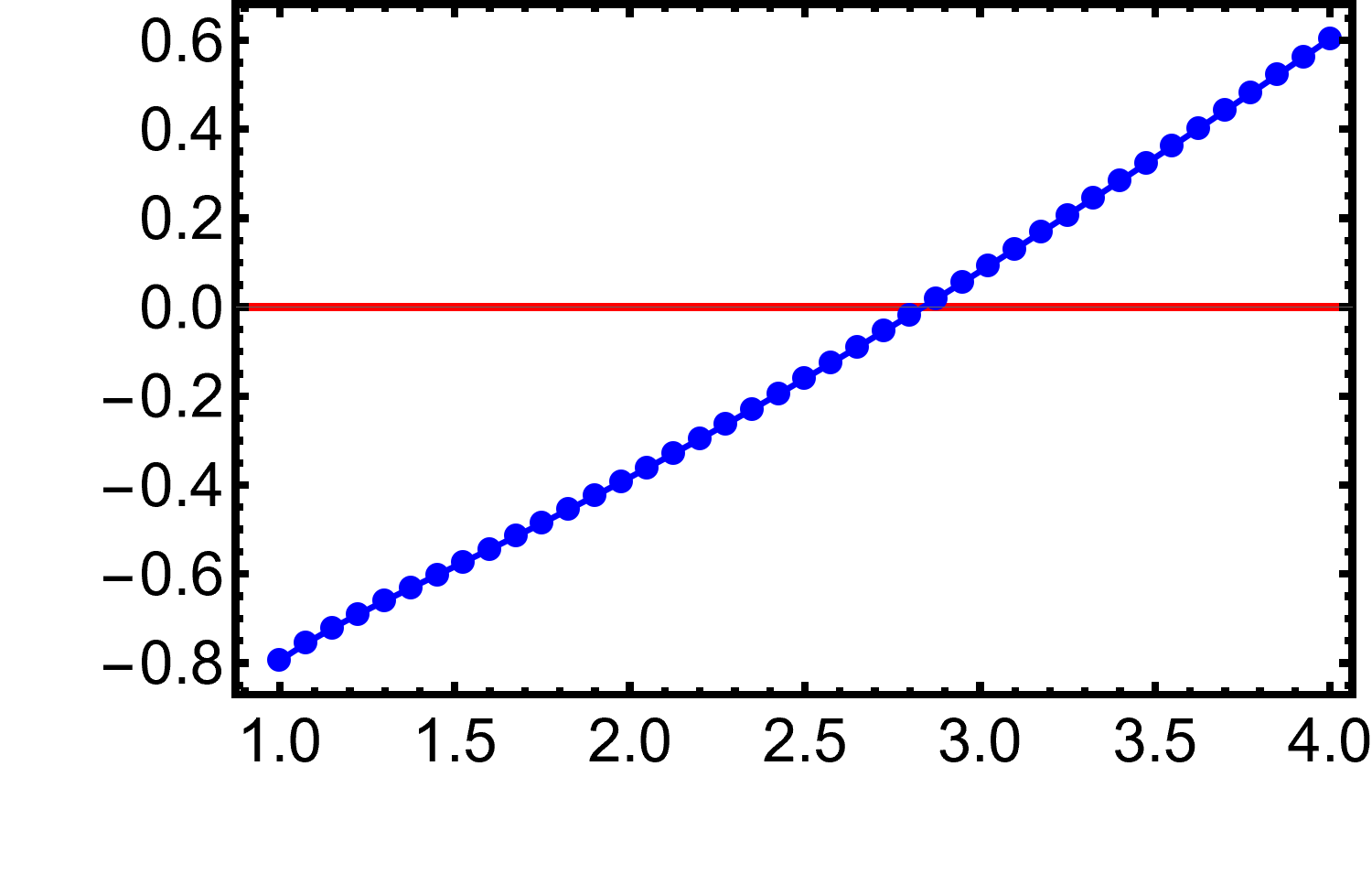}};
  \begin{scope}[x={(img.south east)},y={(img.north west)}]
    \node[fill=white] at (0,0.95) {$\text{(c)}$};
    \node[fill=white] at (0.6,0)  {$ \kb T_{\air }$};
    \node[fill=white] at (0.05,0.65) {$\alpha$};
  \end{scope}
\end{tikzpicture}

\caption{$\alpha$ as a function of $\Delta\mu$, $\chi_{\graze}$ and $ \kb  T_{\air}$. \textcolor{black}{It follows that for low $\kb T_{\air} $ and  $\chi_{\graze}$ as well as high $\Delta \mu$ the coefficient $\alpha$ can flip sign so that a flocking transition can take place.} If not
varied, we took $\Delta \mu = 5$, $\chi_\graze = 0.2$, $\kb T_\air  = 2$, $a = 3$,
$K = 1$ and $m_{\air } = m_\bird  = 1$.}
\label{fig:three-subfigures}
\end{figure}

\begin{figure}[t!]
\centering
\begin{tikzpicture}
  \node[anchor=south west,inner sep=0] (img) at (0,0)
    {\includegraphics[width=0.6\linewidth]{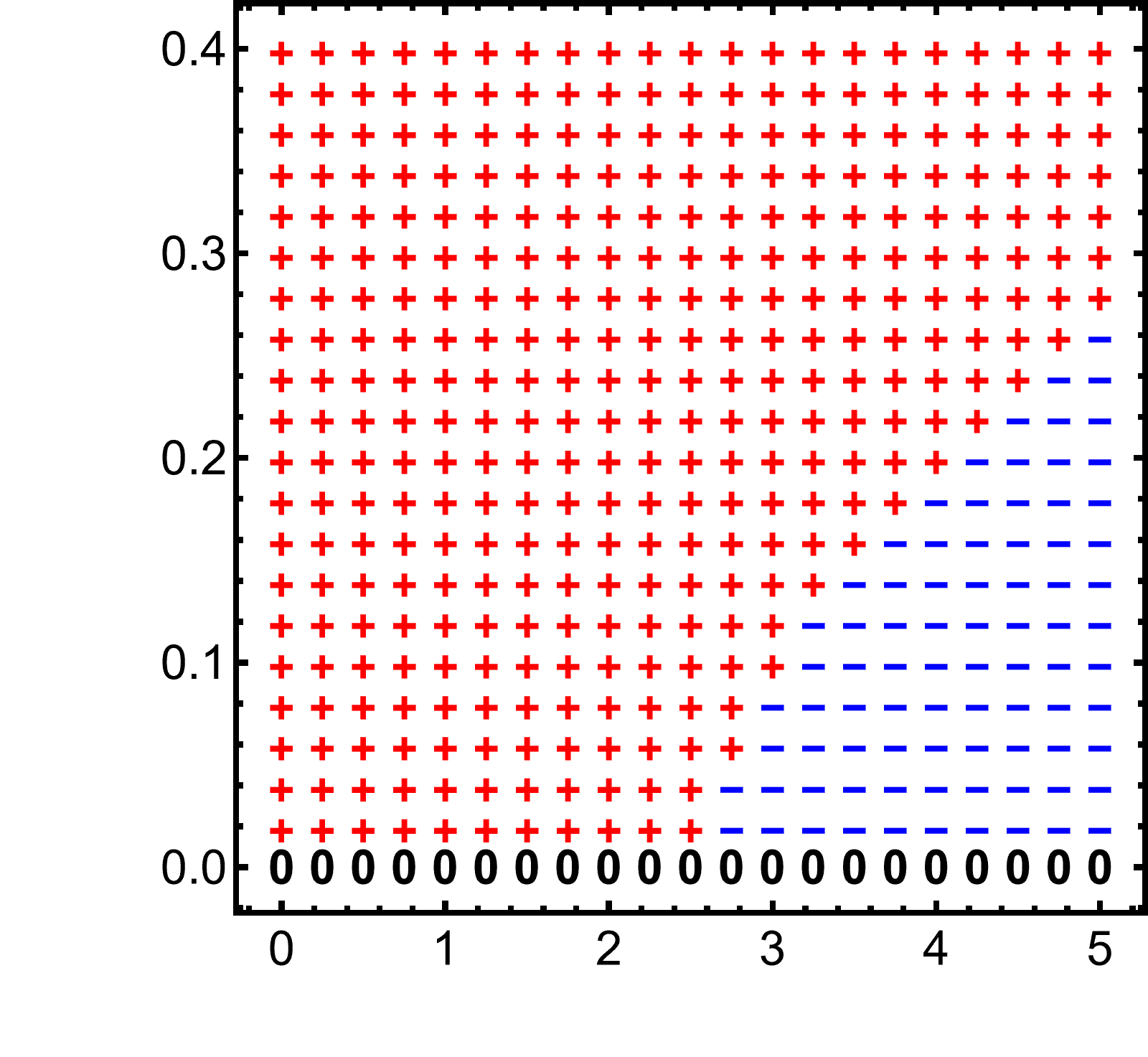}};
  \begin{scope}[x={(img.south east)},y={(img.north west)}]

    \node[fill=white] at (0,0.95) {$\text{(a)}$};
     \node[fill=white] at (0.6,0) {$\Delta \mu$};
    \node[fill=white] at (0,0.6) {$\chi_{\graze }$};

  \end{scope}
\end{tikzpicture}

\vspace{1em}

\begin{tikzpicture}
  \node[anchor=south west,inner sep=0] (img) at (0,0)
    {\includegraphics[width=0.6\linewidth]{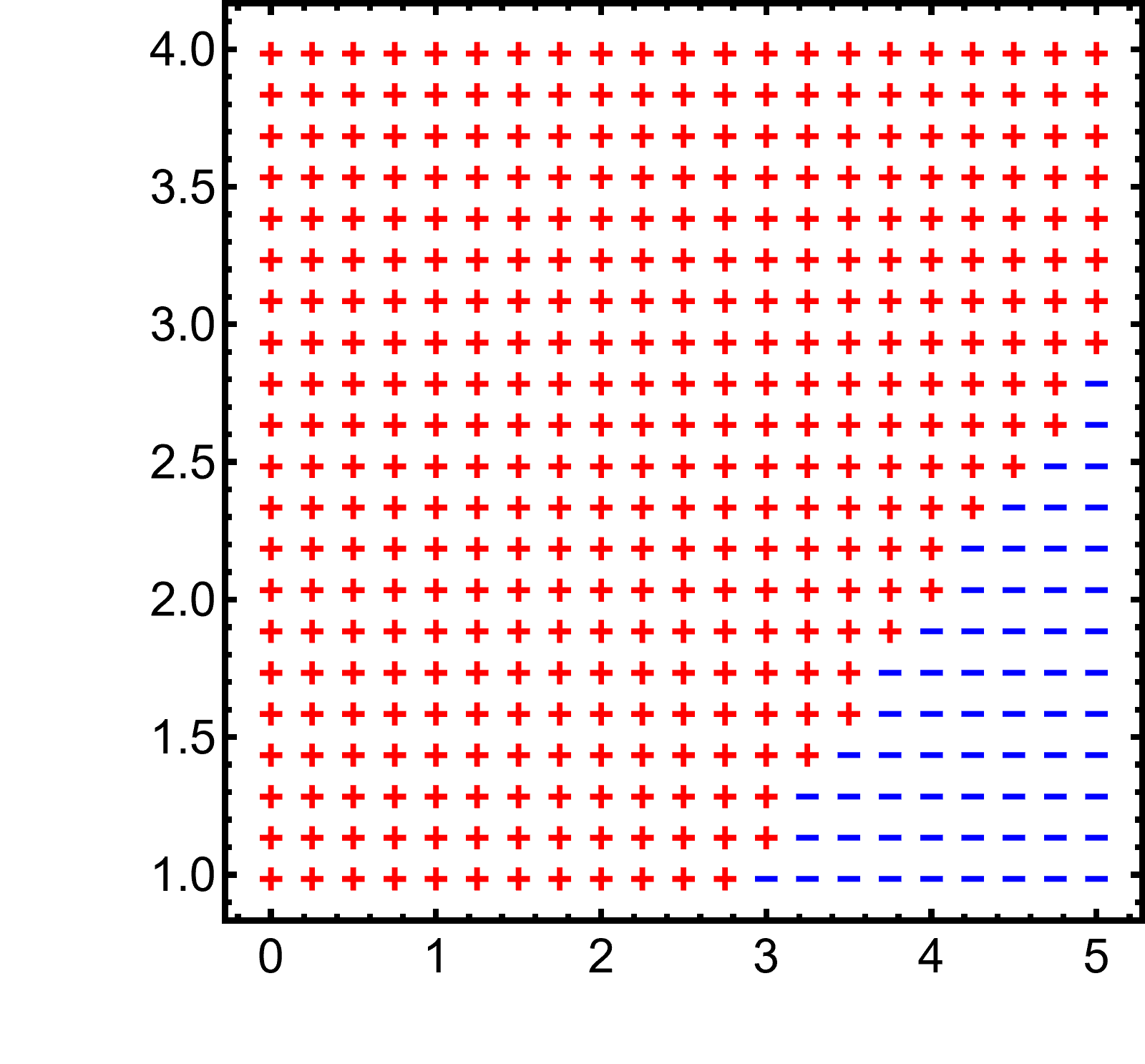}};
  \begin{scope}[x={(img.south east)},y={(img.north west)}]

    \node[fill=white] at (0,0.95) {$\text{(b)}$};
 \node[fill=white] at (0.6,0) {$\Delta \mu $};
    \node[fill=white] at (0,0.6) { $\kb T_{\air}$ };
  \end{scope}
\end{tikzpicture}
\vspace{1em}
\begin{tikzpicture}
  \node[anchor=south west,inner sep=0] (img) at (0,0)
    {\includegraphics[width=0.6\linewidth]{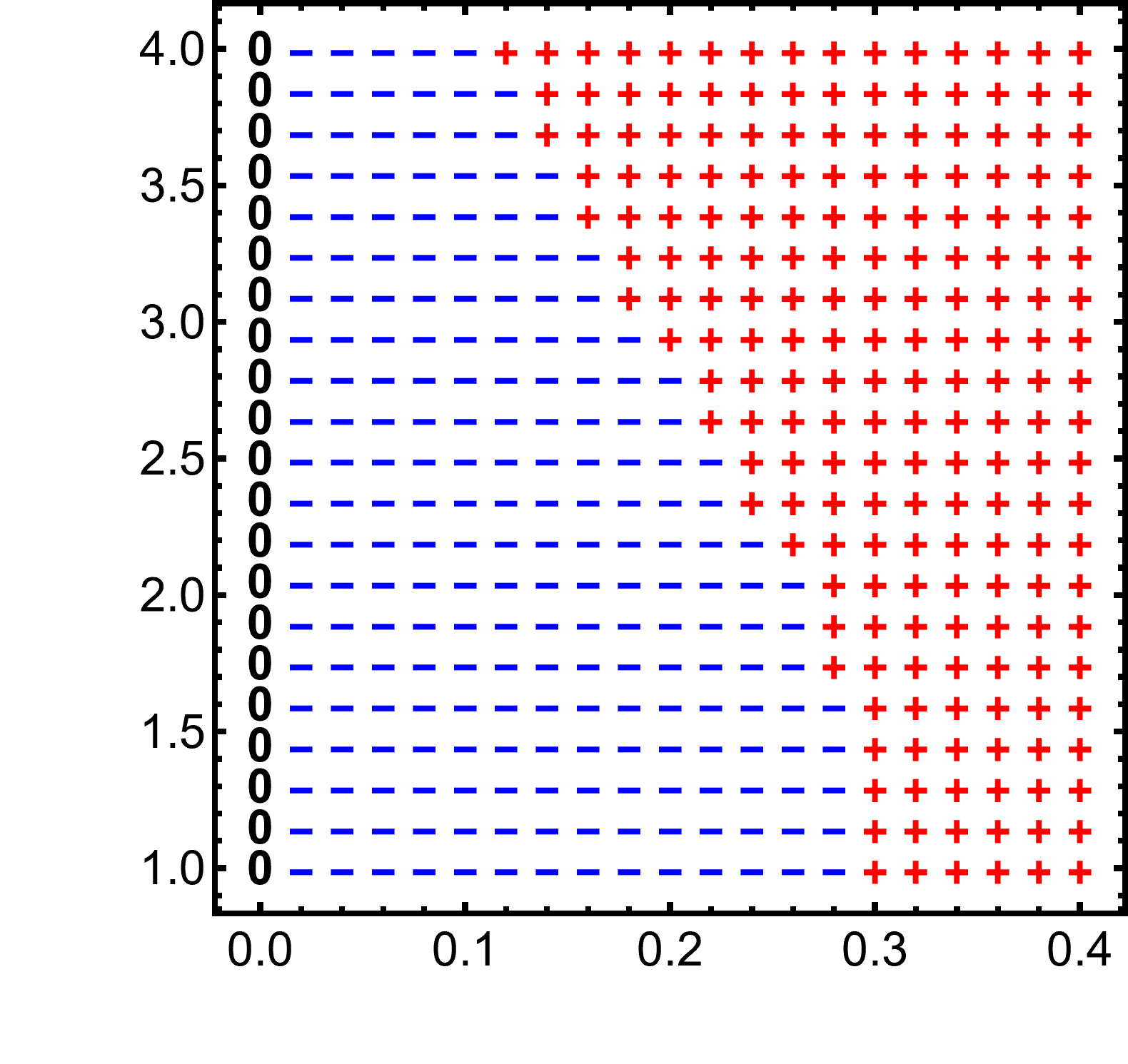}};
  \begin{scope}[x={(img.south east)},y={(img.north west)}]
    \node[fill=white] at (0,0.95) {$\text{(c)}$};
  \node[fill=white] at (0.6,0) {$\chi_{\graze }$};
    \node[fill=white] at (0,0.6) {$\kb T_{\air}$};
  \end{scope}
\end{tikzpicture}

\caption{Diagrams showing the sign of $\alpha$ for different parameter values. \textcolor{black}{Wherever there is a minus sign $\alpha$ no longer corresponds to relaxation but instead gives rise to growth, providing a necessary condition for flocking.} If not
varied, we took the same parameter values as in Fig.~\ref{fig:three-subfigures}.}
\label{fig:three-subfigures123}
\end{figure}

\textcolor{black}{In App.~\ref{app:analyze}, we analyze the integrand of \eqref{eq:flapping} with the upper bound of \eqref{eq:choice} implemented. We see that it in order to make sure that $\alpha \geq 0$, it is important to keep $\chi_{\text{graze}}$ small, as the contribution to the integral from large $\chi$-values is strongly positive. That is, we must filter out large $\chi$ collisions which correspond to frontal bird-air collisions. The requirement that frontal bird-air collisions need to be filtered out is similar to how real birds need to be shaped and positioned in an aerodynamic way in order to attain a nonzero velocity with respect to the ambient air.}
In Fig.~\ref{fig:three-subfigures}, $ \alpha  $ is plotted as a function of $\Delta \mu$, $T$ and $\chi_{\graze }$. We find that increasing $\Delta \mu$ induces flocking, which is anticipated as an increased $\Delta \mu$ \textcolor{black}{quantifies the chemostat and thus the extent to which the system more active. Similarly, it is found that for inducing flocking one must decrease the temperature. This is also expected, as an increased temperature means the birds become more incoherent and thus have more trouble aligning}. The same result is found for numerical simulations of the Vicsek model \cite{Ginelli_2016,Vicsek_1995,Czirók_1997} and for kinetic theory for the Vicsek model \cite{Ihle_2011,bertin2009hydrodynamic} if one identifies the magnitude of noise with temperature. Furthermore, we find that flocking is strongly undermined by mildly increasing $\chi_{\graze}$, from which it follows that frontal bird-air collisions will strongly prevent birds from attaining a nonzero average velocity with air.  
In Fig.~\ref{fig:three-subfigures123}, we provide a diagram that shows the sign of $\alpha$ for the same varied parameters.

\section{Discussion}
In this work, we formulated the simplest possible kinetic theory that is consistent with microscopic reversibility and whose coarse-grained description can display a sign flip of the coefficient $\alpha$ which hints at a flocking transition. The property of microscopic reversibility as well as that the particles do not have a constant velocity magnitude make this model fundamentally different from the seminal Vicsek model as well as any other microscopic model for flocking. In particular, \textcolor{black}{this toy model} treats birds and air as hard spheres which can undergo reactive collisions, so that flocking is emergent and not a consequence of the self-propulsion of the particles considered at microscopic level. \textcolor{black}{For the Vicsek model, interactions between birds are modeled by a velocity orientation that is based on the average velocity orientations of surrounding birds, thereby mimicking real birds that position themselves in a flock using their vision. In this work, aligning interactions between birds simply arise through hard sphere collisions.} 
\newline 
\textcolor{black}{We limited ourselves to computing only one coefficient in the Toner-Tu equations of \eqref{eq:tonertuequations}, which is the coefficient $\alpha $.} The reason why only the coefficient $\alpha $ is computed is twofold.
\begin{enumerate}
    \item To compute the other terms in \eqref{eq:tonertuequations}, in particular the gradient corrections but also  $\beta $, will be a complicated endeavor that likely requires some approximation scheme. 
        \item As was argued in the introduction, it is really the $\alpha$-coefficient that drives the flocking transition, as this is the term which incorporates both the role of the air as well as the activity of bird flocks
    \end{enumerate}
Despite that this work does not achieve the goal of fully deriving all coefficients of \eqref{eq:tonertuequations} from a microscopically reversible model, the microscopic model offers several features which are novel in the context of kinetic theory for flocking: 
\begin{itemize}
  \item An arrow of time arises due to the assumption of molecular chaos
    \item A conserved total energy and momentum exists
    \item $\mathbf{u}_0$ is invariant under Galilean boosts
    \item There exists an $H$-functional whose evolution decays towards zero in the absence of a chemostat
    \item Sufficiently turning on the chemostat allows for velocity growth
    \item The width of the Maxwell-Boltzmann distribution represents noise with the noise magnitude set by temperature
    \item Alignment of bird particles is maintained by nonreactive bird-bird collisions
\end{itemize}
All of these advantages are related to that the microscopic model we formulate is one that is based on passive particles that have the ability to undergo chemical reactions so that a chemostat can be turned on.
\newline 
In addition to the presence of a chemostat and the medium from which birds can draw momentum by flapping their wings, we learned that it is crucial for the flocking transition to occur that the birds are selective with regards to their collisions. To wit, we learned that in order to get a flocking transition, only grazing collisions are allowed, as frontal collisions spoil the formation of a nonzero average bird velocity vector with respect to the average air velocity. \textcolor{black}{The requirement that the birds studied in the present work must be selective about the type of collisions they undergo in order to allow for flocking is similar to how real birds are required to be aerodynamic in order to fly.}
\newline 
In future work, we would like to perform a expansion of the hydrodynamic theory that remains consistent when $\alpha < 0  $. This will be a great challenge for several reasons, one of them being that the Chapman-Enskog expansion relies on the coarse-grained variables to be associated with conservation laws whereas a violation of momentum conservation is inherent to the Toner-Tu equations. One option would be to assume that the theory is still approximately conserved \cite{PhysRevD.108.086011,PhysRevB.107.155108,Hansen2013-qf}, but this assumption certainly no longer holds true when $\alpha < 0 $ as in this case the $\alpha$-coefficient dictates the state at leading order. Another question is whether it makes sense to use the relaxation time approximation \cite{PhysRev.94.511,tong2012kinetic,matus2024nonrelativistictransportframeindifferentkinetic,grosvenor2025hydrodynamicsboostinvariancekinetictheory,PhysRevLett.127.042301} for this endeavor, which has the  potential to dramatically simplify the computation. The problem with the relaxation time approximation in the context of this two-fluid model is that it is the source terms that we are most interested in as they drive the flocking transition. However, the relaxation time approximation will make it so that there is always a source term regardless of the medium. This contribution is typically discarded by hand on the basis that the original collision integral is conserved. This means that when one would compute the source term of the two fluid model one would need to distinguish contributions to the source term that are a result of the bird-air interactions from contributions that come from the relaxation time approximation.
\newline 
Lastly, since the formulation of the collision integral that is formulated in this work is exact, it would be interesting to perform a molecular dynamics simulation \cite{PhysRevB.37.5677,PhysRevA.39.4718,PhysRevLett.52.1333} for two species of hard spheres that undergo the reactive intraspecies collisions described in this work to confirm numerically that the analytically predicted flocking transition occurs.

\section{Acknowledgements}
We thank Aleksander Głódkowski, Kevin Grosvenor, Paweł Matus and Subodh Patil for useful discussions.

\appendix

\onecolumngrid
  
\section{Collision geometry for hard spheres}
\label{eq:measurehardspheres}
\begin{figure}
\centering
\begin{tikzpicture}
  \node[anchor=south west,inner sep=0] (img) at (0,0)
    {\includegraphics[width=0.6\linewidth]{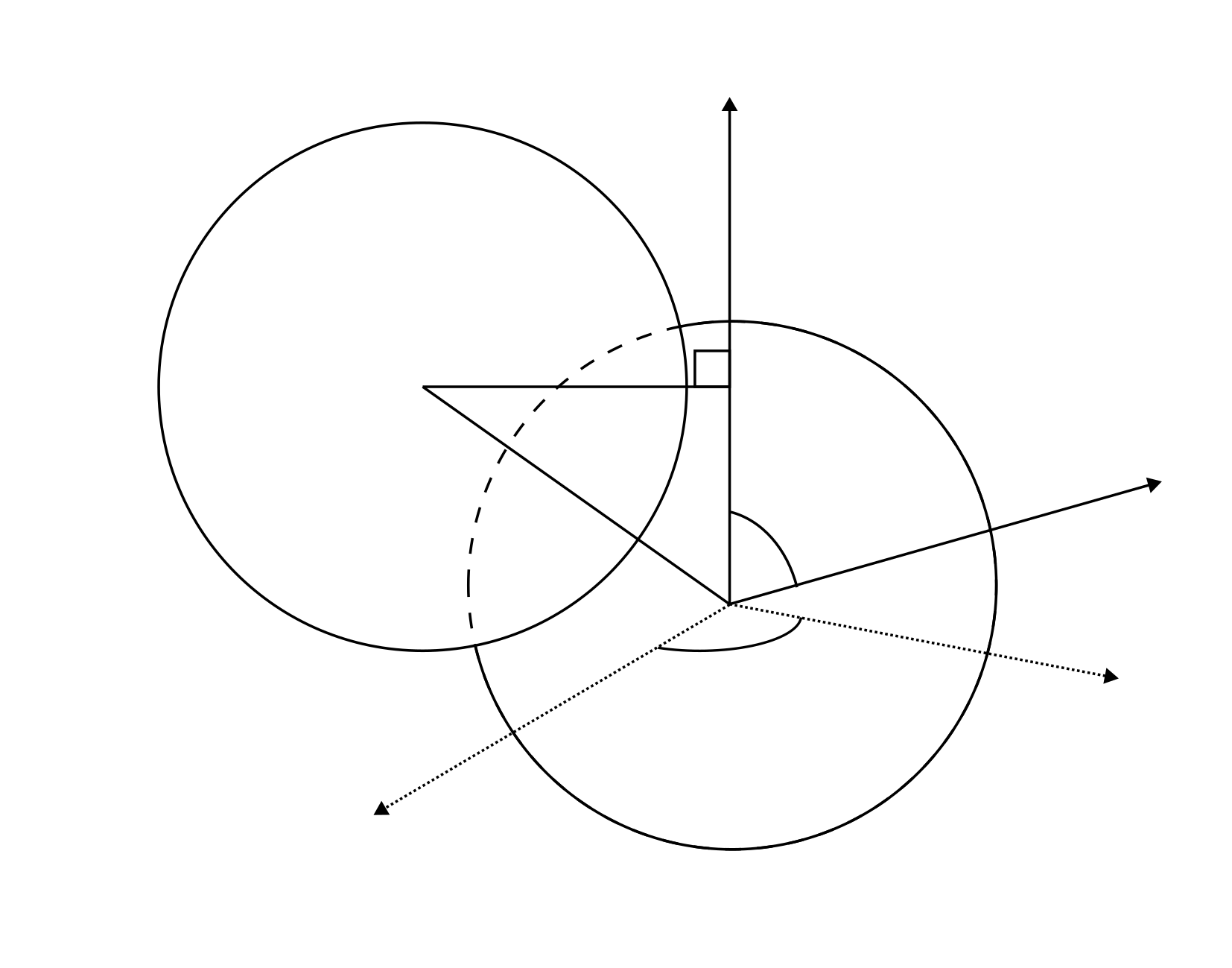}};
  \begin{scope}[x={(img.south east)},y={(img.north west)}, every node/.style={font=\Large}]

    \node[fill=white] at (0.63,0.75) {$\mathbf{e}$};
    \node[fill=white] at (0.83,0.52) {$\mathbf{e}'$};
    \node[fill=white] at (0.67,0.48) {$\chi$};
    \node[fill=white] at (0.45,0.65) {$b$};

  \end{scope}
\end{tikzpicture}
\caption{Three-dimensional depiction of a binary hard-sphere collision.}
\label{figsimpleimpact12}
\end{figure}

In this Appendix we review the basics of the collision geometry of ordinary hard spheres which represent birds. The collision rate per unit volume is given by \cite{chapman1990mathematical,Cercignani1988} (see Fig.~\ref{figsimpleimpact12})
 \begin{align} \label{eq:rateofcollision}
 \mathcal{R}_{\self }  =    \iiint     f_{\bird } ( \bbv )   f_{\bird  } (\bbv_2 )  g b db d \epsilon  d  \bbv  d \bbv_2   ~~ ,  
 \end{align}
where $b$ is the impact parameter of the collision and $\epsilon$ is the azimuthal angle of the collision, which form a cylinder whose length is given by $ g d t $. Let us consider the measure of the orientation $\mep$ of the outgoing relative velocity, which we take to be relative to the $\me$, the orientation of the ingoing relative velocity so that it can be related to the scattering angle $\chi$. It therefore holds that
\begin{align}
    d \mep = \sin(\chi ) d \chi d \epsilon  =  \left( \sin(\chi ) / \left|\frac{\partial b }{ \partial \chi  }  \right| \right)  d b d \epsilon  \equiv  \frac{4}{ d^2_{\bird , \air }  G ( g , \chi ) } b   d b d \epsilon ~~ .  
\end{align}
For nonreactive hard sphere binary collisions, we simply have
\begin{align}  \label{eq:impactparameterformula}
    b =  d_{\bird  } \cos( \chi  / 2 ) ~~ , 
\end{align}
from which it follows that for this simple case it holds that
\begin{align}
    G ( g , \chi )  = \frac{ 4 b  \left|\frac{\partial b }{ \partial \chi  }  \right|  }{d^2_{\bird }  \sin(\chi ) }   = 1   ~~ , 
\end{align}
which means that \eqref{eq:rateofcollision} simplifies to
 \begin{align} \label{eq:rateofcollision1}
   \mathcal{R}_{\self }  =    \iiint     \frac{d^2_{\bird  }}{4}     f_{\bird } ( \bbv  )   f_{\bird } (\bbv_2 )   g   d \mep  d  \bbv  d \bbv_2    ~~  .   
 \end{align}

\section{Scattering angle of reactive collisions}
\label{eq:scatteringangle}
In this Appendix we study the scattering angle for hard sphere particles undergoing a reactive collisions. To compute the scattering angle $\chi$ in relation to the impact parameter $b$ for reactive collisions, we need to account for the fact that the angle between the line of sphere centers and the relative velocity is no longer precisely opposite as is the case in nonreactive collisions. We still however have the relations
\begin{align}  \label{eq:bdrelations}
b^{\pm } =  d_{\bird ,  \air } \sin(\psi^{\pm } )  ~~ , 
\end{align}
and 
\begin{align}  \label{eq:remainingrelation}
\chi^{\pm }  =   \pi - \psi^{\pm } - \psi^{\prime \pm }   ~~ ,   
\end{align}
where $\psi^{\pm}$ is the ingoing angle with respect to the line of sphere centers and $\psi^{\prime \pm}$ is the outgoing angle with respect to the line of sphere centers and $\chi$ is the scattering angle (see Fig.~\ref{figsimpleimpact123}). To obtain $\chi^{\pm }$ one therefore first needs to find how $\psi^{\prime \pm }$ is related to $\psi^{ \pm }$. To find this relation, let us parametrize the relative velocity with respect to the line of sphere centers as
\begin{align}
    \mathbf u^{\prime \pm  }  =   \sqrt{u^2 \pm a^2  }  \{ -  \cos(\psi^{\prime \pm }  ) , \sin(\psi^{\prime \pm }  ) \cos(\phi) , \sin(\psi^{\prime \pm }  ) \sin (\phi)  \}  ~~ , ~~  \mathbf u  =   u \{  \cos(\psi^{ \pm }  ) , \sin(\psi^{ \pm }  ) \cos(\phi) , \sin(\psi^{ \pm }  ) \sin(\phi)   \}  ~~ .  
\end{align}
Where we added a minus for the component of $ \mathbf u^{\prime \pm  }$ longitudinal to the line of sphere centers as this is the component of the relative velocity for which the collision induces a sign flip. Conversely, it follows from the symmetric nature of hard spheres that the transverse part is unchanged, so that we find the relation
\begin{align} \label{eq:psirelation}
  \sin(\psi^{ \prime \pm } )     = \frac{u}{ \sqrt{u^2\pm a^2  }  } \sin(\psi^{\pm} ) ~~ . 
\end{align}
As can also be seen in Fig.~\ref{figsimpleimpact123}, it follows from \eqref{eq:psirelation} that
\begin{align}
    \psi^{  + }   \geq   \psi^{ \prime   + } ~~ , ~~ 
    \psi^{  - }   \leq   \psi^{ \prime   - } ~~ .
\end{align}
Furthermore, from \eqref{eq:psirelation} together with \eqref{eq:bdrelations} it follows that $b^{-}$ is upper bounded as
\begin{align}
    b^{-}  \leq b^-_{\text{max}} ~~ , ~~  b^-_{\text{max}} =  \frac{ \sqrt{u^2 -  a^2  }  }{u} d_{\bird , \air } ~~ , 
\end{align}
whereas $b^+$ can be of any value that is not greater than $d_{\bird , \air }$. Plugging \eqref{eq:psirelation} into \eqref{eq:remainingrelation} leads to
\begin{align}
    \chi^{\pm }  =   \pi - \psi^{\pm } - \sin^{-1} \left( \frac{u}{ \sqrt{u^2 \pm a^2 } }  \sin(\psi^{\pm } )   \right)
\end{align}
We can then write
\begin{align}
\cos(    \chi^{\pm } /2 )   = \sin\left( \frac{1}{2} \psi^{\pm }  + \frac{1}{2} \sin^{-1} \left( \frac{u}{   \sqrt{u^2   \pm a^2  }   }  \sin(\psi^{\pm } )   \right) \right)  ~~ . 
\end{align}
Using \eqref{eq:bdrelations}, we have
\begin{align}  \label{eq:breakdown}
\cos(    \chi^{\pm } /2 )   = \sin\left( \frac{1}{2} \arcsin(\frac{b^{\pm }}{d_{\bird , \air }})  + \frac{1}{2} \arcsin\left(  \frac{u}{   \sqrt{u^2  \pm a^2  }   } \frac{b^{\pm }}{d_{\bird , \air }}  \right) \right)  ~~ . 
\end{align}
 It follows from \eqref{eq:breakdown} that the $\chi^{\pm}$-domain is finite. It holds that
\begin{align}  \label{eq:breakdown123}
 \chi^{\pm }_{\text{min}}  \leq   \chi^{\pm }    \leq  \pi ~~ , 
\end{align}
with
\begin{align}
 \chi^{+  }_{\text{min}} =     2  \arccos\left( \sin\left( \frac{\pi }{4 }     + \frac{1}{2} \arcsin\left(  \frac{u}{   \sqrt{u^2  +  a^2  }   }   \right) \right) \right) =  \arccos\left(  \frac{u}{   \sqrt{u^2  +  a^2  }   }   \right)   ~~ , 
\end{align}
and similarly 
\begin{align}  \label{eq:breakdown12}
    \chi^-_{\text{min}}  =    2 \arccos\left( \sin\left( \frac{1}{2} \arcsin(  \frac{   \sqrt{u^2  -  a^2  }   }{u}  )  + \frac{\pi }{4 } \right) \right) =  \arccos\left(  \frac{ \sqrt{u^2  -  a^2  }}{u}   \right)    ~~ . 
\end{align}
Let us introduce the function $g^{\pm} (u , \chi   )$ defined as
\begin{align} \label{eq:scatteringb}
    b^{\pm } =  d_{\bird , \air }  g^{\pm} (u , \chi   )  ~~ ,   
\end{align}
where we removed the label $\pm$ on the scattering angle $\chi$ as we intend to simultaneously describe collisions of forward and inverse collisions which are tied through microscopic reversibility with a unified scattering angle $\chi $. From \eqref{eq:breakdown} it follows that the equation for $g^{\pm} (u , \chi )$ is given by
\begin{align}  \label{eq:breakdown1}
\cos(    \chi /2 )   = \sin\left( \frac{1}{2} \arcsin(g^{\pm } (u , \chi  )  )  + \frac{1}{2} \arcsin\left(  \frac{u}{   \sqrt{u^2  \pm a^2  }   } g^{\pm } (u , \chi  )   \right) \right)  ~~ . 
\end{align}
When $a$ goes to zero, we simply have
\begin{align}
   \lim_{a \rightarrow 0 }  g^{\pm} (u , \chi  )   =   \cos(\chi  /2 ) ~~ , 
\end{align}
consistent with the expression given in \eqref{eq:impactparameterformula}. One can verify that the solution to \eqref{eq:breakdown1} is given by
\begin{align} \label{eq:trickything2solve12}
\begin{split}
    &  g^{ +   }  (u , \chi   )  =  \frac{\sin(\chi)}{\sqrt{1 +    \frac{u^2 }{  u^2   +   a^2     }  - 2  \frac{u}{   \sqrt{u^2   +   a^2  }   } \cos(\chi) }  }  ,   
\end{split}
\end{align}
Furthermore, it follows from \eqref{eq:breakdown1} that $g^-(u , \chi )$ is related to $g^+(u , \chi )$ as 
\begin{align}  \label{eq:simplification}
   g^- (u , \chi    )  = \Theta (u^2 - a^2) \frac{ \sqrt{u^2 -  a^2 } }{u}   g^+ ( \sqrt{u^2 - a^2 }  ,  \chi     ) ~~ ,  
\end{align}
which will be of use in App.~\ref{app:reactionprob}.

\section{Microscopically reversible collision integral}
\label{app:reactionprob}
In this Appendix we impose microscopic reversibility for the reactive volume $\kappa^{\pm} (u , \chi )$. Let us consider the collision rate per unit volume for the forward and inverse reactive collision, which similar to App.~\ref{eq:measurehardspheres} is given by
\begin{align} \label{eq:rateofcollision1212}
 \mathcal{R}^{\pm}_{\flap }  =    \rho^{\pm}  \iiint     \kappa^{\pm }  ( u , \chi ) f_{\bird } ( \bbv )   f_{\bird  } (\bbw_2 )  u b db d \epsilon  d  \bbv  d \bbw_2   ~~ ,  
 \end{align}
 where we added a factor $\kappa^{\pm }  ( u , \chi ) \rho^{\pm} $ to account for our assumptions that bird-air collisions only happen with a probability given by \eqref{eq:condprob}. Using App.~\ref{eq:scatteringangle}, we can substitute $b$ for $\chi$, leading to
 \begin{align}
 \mathcal{R}^{\pm}_{\flap }  =  \frac{d^2_{\bird , \air }}{4}   \rho^{\pm}  \iiint  d \mathbf{v} d  \mathbf{w}_2  d \mathbf{e}^{ \prime  }    \,      G^{\pm}  (u , \chi   )   u   \kappa^{\pm}  ( u , \chi   )           f_{\bird} ( \mathbf{v} ) f_{\air} (  \mathbf{w}_2    )      
 \end{align}
 where we introduced the function
\begin{align}  \label{eq:Gformula}
    G  ^{ \pm  } (u , \chi   )  & =  4 g^{ \pm }  (u , \chi   )   \left( \left|\frac{\partial g^{ \pm  }  (u , \chi   )   }{ \partial \chi  }  \right| / \sin(\chi )  \right)    ~~ . 
\end{align}
 Let us then use the relations
   \begin{align} \label{eq:uequations123123}
 \mathbf{v}    = \mathbf{U} + \frac{m_{\air}}{m_0}  \mathbf{u}  ~~ ,  ~~         \mathbf{w}_2     = \mathbf{U} - \frac{m_{\bird}}{m_0}  \mathbf{u}   ~~  ,  \end{align}
 where $\mathbf{U}$ is the center of mass velocity. Rewriting the integral of \eqref{eq:rateofcollision} in terms of $\mathbf{u}$ and $\mathbf{U}$ comes with a unit Jacobian determinant \cite{chapman1990mathematical}, so that we have
\begin{align} \label{eq:rateofcollision12}
 \mathcal{R}^{\pm}_{\flap }  =     \frac{d^2_{\bird , \air }}{4}   \rho^{\pm}  \iiint  d  \mathbf{U}  d \mathbf{u}  d \mathbf{e}^{ \prime  }    \,      G^{\pm}  (u , \chi   )   u   \kappa^{\pm}  ( u , \chi   )           f_{\bird} ( \mathbf{v} ) f_{\air} (  \mathbf{w}_2    )     ~~ ,  
 \end{align}
 To see how these reactive collisions could be microscopically reversible, let us now consider the rate at which collisions inverse to \eqref{eq:rateofcollision12} take place. To wit, for every collision described by $u$ and $\chi$, there is an equivalent reverse collision given by $u^{\pm} $ and $ \chi$, where
 \begin{align}
     u^{\pm} = \sqrt{u^2 - a^2 } ~~ . 
 \end{align}
Let us start from the rate of reversed collisions 
\begin{align} \label{eq:rateofcollision12131xxx3}
 \mathcal{R}^{\pm , \text{rev} }_{\flap }  =    \rho^{\mp}  \iiint   b db d \epsilon  d  \bbv_1^{\prime \mp }  d \bbw^{\prime \mp }_2  u^{\mp }  \kappa^{\mp }  ( u^{\mp } , \chi ) f_{\bird } ( \bbv_1^{\prime \mp } )   f_{\bird  } (\bbw_2^{\prime \mp } )    ~~ .  
 \end{align}
 Following the same steps as above, \eqref{eq:rateofcollision12131xxx3} can be written as
 \begin{align} \label{eq:rateofcollision1213123zz}
 \mathcal{R}^{\pm , \text{rev} }_{\flap }  =  \frac{d^2_{\bird , \air }}{4}  \rho^{\mp}  \iiint   d \mathbf{u}^{\prime \mp }  d \mathbf{U} d \mathbf{e}^{\prime }   G^{\mp}  (u^{\mp} , \chi   )  u^{\mp }  \kappa^{\mp }  ( u^{ \mp} , \chi ) f_{\bird } ( \bbv_1^{\prime \mp } )   f_{\bird  } (\bbw_2^{\prime \mp } )    ~~ ,  
 \end{align}
 where we note that we did not give $\mathbf{U}$ a label as the center of mass velocity is a collisional invariant even for reactive collisions. To make sure that \eqref{eq:rateofcollision1213123zz} is on equal footing with \eqref{eq:rateofcollision12}, we must change the integration variable $u^{\pm}$. We then use that \cite{10.1063/1.1725565,10.1063/1.1731889,kugerl,LIGHT1969281}
\begin{align}  \label{eq:measurething}
u^{  \pm} d  \mathbf{u}^{\prime  \pm} d  \mathbf{e}   =  u ( u^{2 } \mp a^2  ) d  u  d  \mathbf{e}^{\prime }  d  \mathbf{e}    =  \frac{ ( u^{2 } \mp a^2  )  }{u  }  d  \mathbf{u} d  \mathbf{e}^{\prime }  
\end{align}
as well as \eqref{eq:simplification} from which together with \eqref{eq:Gformula} it follows that 
\begin{align}  \label{eq:Gthing123123}
  ( u^{2 } \mp a^2  )  G^{\pm} (\sqrt{ u^{2 } \mp a^2  }   , \chi   ) =   u^2    G^{ \mp } (u , \chi   ) ~~ ,  
\end{align}
which is \eqref{eq:Gthing}. Combining \eqref{eq:Gthing123123} and \eqref{eq:measurething} gives
\begin{align}  \label{eq:Gidentity}
  u^{  \pm} G^{\pm} ( u^{  \pm}   , \chi   )   d  \mathbf{u}^{\prime  \pm} d  \mathbf{e}    =  u G^{ \mp } (u , \chi   ) d  \mathbf{u} d  \mathbf{e}^{\prime  }  ~~ . 
\end{align} 
\eqref{eq:rateofcollision1213123zz} thus turns into 
\begin{align} \label{eq:rateofcollision121312dd3}
 \mathcal{R}^{\pm , \text{rev} }_{\flap }  =  \frac{d^2_{\bird , \air }}{4}  \rho^{\mp}  \iiint   d \mathbf{u}  d \mathbf{U} d \mathbf{e}^{\prime }   u   G^{ \pm } (u , \chi   ) \kappa^{\mp }  ( \sqrt{ u^2  \pm a^2  } , \chi ) f_{\bird } ( \bbv_1^{\prime \mp } )   f_{\bird  } (\bbw_2^{\prime \mp } )    ~~ ,  
 \end{align}
Let us now compare \eqref{eq:rateofcollision121312dd3} to \eqref{eq:rateofcollision12}. We see that both collision rates are proportional to a bird and air density as well as well as the reactive volume $ \kappa^{\pm }  ( \sqrt{ u^2  \pm a^2  } , \chi ) $, which acts as a scattering function of a three-body collision. To make sure that this three-body collision occurs in a way where there is no distinction between the forward and reverse collision, we impose 
\begin{align} \label{eq:kapprelations}
     \kappa^{- } (u ,\chi )      =   \Theta (u^2  - a^2  )\kappa^{+} (\sqrt{u^2  - a^2} , \chi  )   ~~ .    
\end{align}
It follows then that collision integral can be written as
 \begin{align}  \label{eq:frict1233}
\mathcal{C}^{\air}_{\flap    }  =    \frac{d^2_{\bird , \air }}{4}   \sum_{\pm } \iint  d  \mathbf{w}_2   d \mathbf{e}'   u   \,   \kappa^\mp (u , \chi  )  G^{ \mp } (u , \chi   )     \left( \rho^{\pm }   f_{\bird} (\mathbf{v}^{\prime \pm }_1 ) f_{\air} (  \mathbf{w}^{\prime \pm  }_2   )      - \rho^{\mp }             f_{\bird} ( \mathbf{v} ) f_{\air} (  \mathbf{w}_2    )    \right)     &      ~~ . \end{align}
Lastly, since $G^{  \pm    } (u , \chi   )$ originates from a Jacobian determinant, it follows that $G^{\pm} (u , \chi )$ obeys the simple integral relations
\begin{subequations}  \label{eq:usefulidentities}
        \begin{align}
    \int_0^{d_{\bird , \air }} b d b   &   =  \frac{d^2_{\bird , \air }}{4}   \int^{\pi }_{\chi^+_{\min} } d \chi  \sin(\chi )  G^{  +  } ( u,\chi   ) = \frac{d^2_{\bird , \air }}{2}    ~~ ,   \\ 
     \int_0^{b^-_{\max}} b d b   &  =  \frac{d^2_{\bird , \air }}{4}  \int^{\pi }_{\chi^-_{\min} }  d \chi    \sin(\chi ) G^{- } ( u,\chi   )  =    \frac{d^2_{\bird , \air }}{2}  \frac{  u^2 - a^2   }{u^2  }  ~~  .  
\end{align}
\end{subequations}

\section{Total energy conservation of flapping collisions}
\label{app:energyproof}
In this Appendix we show that the energy conservation constraint on the collision integral given by \eqref{eq:energyconservation111} is obeyed. We start from
\begin{align}
\begin{split}  \label{eq:Htheoremthing12}
   & \sum_{\pm}  \left(  \frac{1}{2}   \int d \mathbf{v }_1  m_\bird v_1^2 \mathcal{C}^{\bird \pm  }_{\flap }   +  \frac{1}{2}  \int d \mathbf{w}_1     m_\air w_2^2 \mathcal{C}^{\air \pm  }_{\flap }  \right)  
  =   \\  &   \sum_{\pm }\frac{  d^2_{\bird, \air}}{4}   \iint  d\mathbf{v}_1   d  \mathbf{w}_2    d \mathbf{e}' \,      u   G^{\pm } ( u  ,   \chi  )      \kappa^{\pm }_{\flap } \left( u  , \chi   \right)    \left(    \frac{1}{2} m_\bird v^2_1 +   \frac{1}{2} m_{\air } w^2_2    \right)   \\  & 
 \cdot \left(   \rho^{\mp  }   
     f_{\bird} ( \mathbf{v}^{\mp  \prime}_1 , t) f_{\air} (  \mathbf{w}^{\mp  \prime}_2  , t  ) -      \rho^{\pm   }     f_{\bird} (\mathbf{v}_{1 } , t) f_{\air} (  \mathbf{w}_{2 }  , t  )    \right)  ~~   , 
     \end{split}
\end{align}
Using \eqref{eq:alphaterms}, \eqref{eq:energyconservation123} and \eqref{eq:Gidentity} we can rewrite \eqref{eq:Htheoremthing12} as 
\begin{align}  \label{eq:ffsfmlxx}
\begin{split}
       ... =  & \frac{  d^2_{\bird, \air}}{4}  \iiint  d\mathbf{v}_1   d  \mathbf{w}_2    d \mathbf{e}' \,       u   G^{ +  } ( u  , \chi  )    \kappa^{+}_{\flap } \left( u  ,\chi   \right)     \left(  \frac{1}{2} m_\bird v^2_1 +  \frac{1}{2} m_{\air } w^2_2    \right)     \left(   \rho^{-  }   
     f_{\bird} ( \mathbf{v}^{- \prime}_1 ) f_{\air} (  \mathbf{w}^{- \prime}_2    ) -      \rho^{+  }     f_{\bird} (\mathbf{v}_{1 }) f_{\air} (  \mathbf{w}_{2 }    )    \right)    \\ 
 & +\frac{  d^2_{\bird, \air}}{4}    \iiint  d\mathbf{v}^{+ \prime }_1   d  \mathbf{w}^{+ \prime }_2    d \mathbf{e} \,   u^{+ }    G^{ +  } ( u^{+ } , b )     \kappa^{+}_{\flap } \left( u^{+ }  ,\chi      \right)    \left(  \frac{1}{2}  m_\bird  ( v_1^{ \prime +  } )^2+ \frac{1}{2}  m_{\air }  (   w_2^{ \prime +  } )^2  +  \Delta E      \right)    \\  &   \cdot   \left(   \rho^{+  }   f_{\bird} ( \mathbf{v}^{+ \prime}_1 ) f_{\air} (  \mathbf{w}^{+ \prime}_2    ) -      \rho^{-  }     f_{\bird} (\mathbf{v}_{1 }) f_{\air} (  \mathbf{w}_{2 }    )    \right)   \,  . 
     \end{split}
\end{align}
Since $\mathbf{v}^{+ \prime }_1 $ and $  \mathbf{w}^{+ \prime }_2 $ are dummy variables, they can be renamed. We must however take note of how they are tied to $\mathbf{v}_1  $ and $  \mathbf{w}_2 $. To account for this, we realize that the outgoing velocities are tied energetically to ingoing forward reactive velocities the same way that ingoing velocities are to outgoing inverse reactive velocities, which justifies the reparametrization
\begin{align}
    \mathbf{v}^{+ \prime }_1   ,   \mathbf{w}^{+ \prime }_2  \rightarrow   \mathbf{v}_1   ,   \mathbf{w}_2  ~~ , ~~  \mathbf{v}_1   ,   \mathbf{w}_2 \rightarrow  \mathbf{v}^{- \prime }_1   ,    \mathbf{w}^{- \prime }_2
\end{align}
  so that \eqref{eq:ffsfmlxx} reduces to
\begin{align}  \label{eq:ffsfml123123}
\begin{split}
      & ...  = - \frac{  d^2_{\bird, \air}}{4}  \Delta E   \iiint  d\mathbf{v}_1   d  \mathbf{w}_2    d \mathbf{e}' \,       u    G^{+}  \left( u , \chi \right)   \kappa^{+}_{\flap } \left( u  , \chi  \right)       \left(   \rho^{-  }   
     f_{\bird} ( \mathbf{v}^{- \prime}_1 ) f_{\air} (  \mathbf{w}^{-  \prime}_2    ) -      \rho^{+  }     f_{\bird} (\mathbf{v}_{1 }) f_{\air} (  \mathbf{w}_{2 }    )    \right)  = -  \Delta E  \,   \mathcal{C}^{\text{C}   +       }_{\flap }  \,  . 
     \end{split}
\end{align}
We can thus conclude, using \eqref{eq:plusminusrelation}
\begin{align}
 \sum_{\pm}  \left(   \frac{1}{2}   \int d \mathbf{v }_1  m_\bird v_1^2 \mathcal{C}^{\bird \pm  }_{\flap }   +   \frac{1}{2}  \int d \mathbf{w}_2     m_\air w_2^2 \mathcal{C}^{\air \pm  }_{\flap }  \right)  
  =    \Delta E  \,   \mathcal{C}^{\text{C}   -     }_{\flap } ~~ , 
\end{align}
which is \eqref{eq:energyconservation111}.

\section{$H$-theorem}
\label{app:Htheorem}
In this section we show that
\begin{align}
    \frac{\partial H}{ \partial t} \leq 0 ~~ . 
\end{align}
In particular, we show this for $ H_{\flap}$, as this is the only part of $H$ for which the nonpositivity of the time evolution is not textbook knowledge \cite{chapman1990mathematical,Cercignani1988,tong2012kinetic}. Plugging \eqref{eq:flapflapcoll} into \eqref{eq:flapequation}, we find
\begin{align}  \label{eq:Htheoremthing}
\begin{split}
       \frac{\partial H_{\flap}}{ \partial t} = &  \sum_{\pm }\frac{  d^2_{\bird, \air}}{4}   \iiiint d \mathbf{r}  d\mathbf{v}_1   d  \mathbf{w}_2    d \mathbf{e}' \,      u   G^{\pm } ( u  ,   \chi  )      \kappa^{\pm }_{\flap } \left( u , \chi   \right)     \log (  \rho^{\pm  } f_{\bird}(\mathbf{v}_1 )  f_{\air }(\mathbf{w}_2) )    \\  & 
 \cdot \left(   \rho^{\mp  }   
     f_{\bird} ( \mathbf{v}^{\mp  \prime}_1 ) f_{\air} (  \mathbf{w}^{\mp  \prime}_2    ) -      \rho^{\pm   }     f_{\bird} (\mathbf{v}_{1 } ) f_{\air} (  \mathbf{w}_{2 }    )    \right)    , 
     \end{split}
\end{align}
where we used that particle number conservation guarantees that
\begin{align}
\begin{split}
 \sum_{\pm } \iiiint d \mathbf{r}  d\mathbf{v}_1   d  \mathbf{w}_2   d \mathbf{e}' \,       u   G^{\pm } ( u , \chi   )   \kappa^{\pm}_{\flap } \left( u , \chi   \right)     \left(   \rho^{\mp  }   
     f_{\bird} ( \mathbf{v}^{\mp \prime}_1 ) f_{\air} (  \mathbf{w}^{\mp \prime}_2    ) -      \rho^{\pm  }     f_{\bird} (\mathbf{v}_{1 }) f_{\air} (  \mathbf{w}_{2 }    )    \right)   =0 ~~   , 
     \end{split}
\end{align}
as well as \eqref{eq:plusminusrelation}. In \eqref{eq:Htheoremthing} we ignored possible spatial dependence of the $H$-functional, because spatial dependence which only gives rise to a spatial total derivative term which vanishes when integrating the $H$-functional over all space \cite{tong2012kinetic}. Now, using \eqref{eq:alphaterms} as well as \eqref{eq:Gidentity}, we can write \eqref{eq:Htheoremthing} as 
\begin{align}  \label{eq:ffsfml}
\begin{split}
      & \frac{\partial H_{\flap}}{ \partial t} =\frac{  d^2_{\bird, \air}}{4}  \iiiint d \mathbf{r}  d\mathbf{v}_1   d  \mathbf{w}_2    d \mathbf{e}' \,       u   G^{ +  } ( u  , \chi  )    \kappa^{+}_{\flap } \left( u    , \chi  \right)      \log (\rho^{+ }  f_{\bird}(\mathbf{v}_1 )  f_{\air }(\mathbf{w}_2) )       \left(   \rho^{-  }   
     f_{\bird} ( \mathbf{v}^{- \prime}_1 ) f_{\air} (  \mathbf{w}^{- \prime}_2    ) -      \rho^{+  }     f_{\bird} (\mathbf{v}_{1 }) f_{\air} (  \mathbf{w}_{2 }    )    \right)    \\ 
 & +\frac{  d^2_{\bird, \air}}{4}    \iiiint d \mathbf{r} d\mathbf{v}^{+ \prime }_1   d  \mathbf{w}^{+ \prime }_2    d \mathbf{e} \,   u^{+ }    G^{ +  } ( u^{+ } , \chi  )     \kappa^{+}_{\flap } \left( u^{+ }    , \chi   \right)     \log ( \rho^{- }  f_{\bird}(\mathbf{v}_1 )  f_{\air }(\mathbf{w}_2) )       \left(   \rho^{+  }   
     f_{\bird} ( \mathbf{v}^{+ \prime}_1 ) f_{\air} (  \mathbf{w}^{+ \prime}_2    ) -      \rho^{-  }     f_{\bird} (\mathbf{v}_{1 }) f_{\air} (  \mathbf{w}_{2 }    )    \right)   \,  . 
     \end{split}
\end{align}
Since $\mathbf{v}^{+ \prime }_1 $ and $  \mathbf{w}^{+ \prime }_2 $ are dummy variables, they can be renamed, however we must take note of how they are tied to $\mathbf{v}_1  $ and $  \mathbf{w}_2 $. To account for this, we realize that the outgoing velocities are tied energetically to ingoing forward reactive velocities the same way that ingoing velocities are tied to outgoing inverse reactive velocities, which justifies the reparametrization
\begin{align}
    \mathbf{v}^{+ \prime }_1   ,   \mathbf{w}^{+ \prime }_2  \rightarrow   \mathbf{v}_1   ,   \mathbf{w}_2  ~~ , ~~  \mathbf{v}_1   ,   \mathbf{w}_2 \rightarrow  \mathbf{v}^{- \prime }_1   ,    \mathbf{w}^{- \prime }_2
\end{align}
  so that \eqref{eq:ffsfml} can be rewritten as
\begin{align}  \label{eq:ffsfml123}
\begin{split}
       \frac{\partial H_{\flap}}{ \partial t} = & \frac{  d^2_{\bird, \air}}{4}  \iiiint d \mathbf{r}  d\mathbf{v}_1   d  \mathbf{w}_2    d \mathbf{e}' \,       u    G^{+}  \left( u , \chi \right)   \kappa^{+}_{\flap } \left( u  , \chi  \right)     \log ( \frac{ \rho^+   f_{\bird}(\mathbf{v}_1 )  f_{\air }(\mathbf{w}_2) }{ 
 \rho^- f_{\bird}(\mathbf{v}^{- \prime }_1 )  f_{\air }(\mathbf{w}^{- \prime }_2) } )     \\  &  \cdot   \left(   \rho^{-  }   
     f_{\bird} ( \mathbf{v}^{- \prime}_1 ) f_{\air} (  \mathbf{w}^{-  \prime}_2    ) -      \rho^{+  }     f_{\bird} (\mathbf{v}_{1 }) f_{\air} (  \mathbf{w}_{2 }    )    \right)   \,  . 
     \end{split}
\end{align}
Since the integrand is of the form $\log(x/y) ( y -x  )  $ and it holds that
\begin{align}
    \log(x/y) ( y -x  ) \leq 0  ~~ , ~~ x , y >  0  ~~ , 
\end{align}
we find the inequality 
\begin{align}
     \frac{\partial H_{\flap}}{ \partial t}  \leq  0 ~~ . 
\end{align}

\section{Growth rate}
\label{app:sourceterm}
Taking the collision integral of \eqref{eq:frictionaxxxxx6123} we can construct the source term of \eqref{eq:sourceterm} which leads us to
\begin{align}  \label{eq:intermediatestep}
      \begin{split}
    \mathbf{S}  =  &   \frac{d^2_{\bird , \air }}{4 n_{\bird }}  \sum_{\pm}  \rho^{\mp}  \iiint d    \mathbf{U}_0    d    \mathbf{u}  d \mep u G^{\mp} (u , \chi)  \kappa^{\mp} (u , \chi )  \mathbf{V}             f^{(0)}_{\bird} (\mathbf{v})  f^{(0)}_{\air}  (\mathbf{w}_2 ) \left( \exp\left( 
  \beta_{\air } \left(  \mur    
    (  \mathbf{u}^{\prime  \pm  }  -  \mathbf{u} )   \cdot \mathbf{u}_0  
 \mp \Delta \mu \right)  \right)  -  1  \right)       ~~  ,     \end{split}  
 \end{align}
 where 
  \begin{align} \label{eq:importantffterm}
 \begin{split}
      &    f^{(0)}_{\bird} (\mathbf{v})  f^{(0)}_{\air}  (\mathbf{w}_2 )  =  n_{\bird } n_{\air } \left(  \frac{\sqrt{m_{\bird } m_{\air } } }{2 \pi \kb  T_{\air  } } \right)^{3}   \exp( - \beta_{\air }  \left(  \frac{1}{2}( m_{\bird} V^{ 2}  + m_{\air} W^{ 2}_2  )   \right) )  .  \end{split}
 \end{align}
 and we used that because of \eqref{eq:equationflap} we have
 \begin{align}
 \int d \mathbf{v}  \mathbf{v}\mathcal{C}_{\flap }  =   \int d \mathbf{v}  \mathbf{V}\mathcal{C}_{\flap }    +\mathbf{v}_0  \int d \mathbf{v}  \mathcal{C}_{\flap }   =  \int d \mathbf{v}  \mathbf{V}\mathcal{C}_{\flap }    ~~ . 
 \end{align}
 Let us now introduce the center of mass velocity which together with relative velocity $\mathbf{u}$ is related to the bird and air velocities as
    \begin{align} \label{eq:uequations}
 \mathbf{v}    = \mathbf{U} + \frac{m_{\air}}{m_0}  \mathbf{u}  ~~ , ~~     \mathbf{w}_2   = \mathbf{U} - \frac{m_{\bird}}{m_0}  \mathbf{u}   ~~ .   \end{align}
        It holds that
        \begin{align}
        \frac{1}{2}  \left( m_{\bird}   v^2 +   m_{\air} w_2^2  \right)   =  \frac{1}{2}  \left(m_0  U_0^2   +  \mur   u^{ 2 } \right)  ~~ . 
        \end{align}
        Let us then define
        \begin{align}
             \mathbf{U}_0 =  \mathbf{U} - \mathbf{v}_0 ~~, 
        \end{align}
so that \eqref{eq:uequations} can be written as        
    \begin{align} \label{eq:uequations123}
 \mathbf{V}    = \mathbf{U}_0  + \frac{m_{\air}}{m_0}  \mathbf{u}  ~~ , ~~      \mathbf{W}_2    = \mathbf{U}_0 + \mathbf  u_0  - \frac{m_{\bird}}{m_0}  \mathbf{u}  ~~ .   \end{align}
        For the Maxwell-Boltzmann distributions of \eqref{eq:importantffterm} we can then substitute \eqref{eq:uequations123} according to
 \begin{align}  \label{eq:energybalance}
             \frac{1}{2}( m_{\bird} V^{ 2}  + m_{\air} W^{ 2}_2  )  = \frac{1}{2}  (m_0  U_0^2   +  \mur   u^{ 2 }  +  m_{\air} u^2_0   )  + m_{\air }  \mathbf{u}_0 \cdot \mathbf{U}_0  -  \mur   \mathbf{u}_0 \cdot \mathbf{u}  ~~ . 
        \end{align}
With these integration variables we find the first order contribution
  \begin{align}  \label{eq:intermediatestep11}
      \begin{split}
  &  \mathbf{S}^{[1]}    =  \frac{ \mur  \beta_{\air }  m_{\air }}{m_0 }  \frac{d^2_{\bird , \air }}{4 n_{\bird }} \sum_{\pm}   \rho^{\mp}    \iiint  d  \mathbf{u} d  \mathbf{U}_0  d \mep   \,   u    \mathbf u    [ f^{(0)}_{\bird}  f^{(0)}_{\air}  ]_0     G^{\mp} (u , \chi   ) \kappa^{\mp}_{\flap} (u  , \chi  )    
    ( \mathbf{u}^{ \pm  \prime }  -  \mathbf{u} )   \cdot \mathbf{u}_0  
\exp(  \pm  \beta_{\air } \Delta \mu   )    \\ 
   & -    \frac{\beta_{\air }  m_{\air }}{m_0 } \frac{d^2_{\bird , \air }}{4 n_{\bird }}  \sum_{\pm}  \rho^{\mp}     \iiint  d  \mathbf{u} d  \mathbf{U}_0  d \mep   \,   u       [ f^{(0)}_{\bird}  f^{(0)}_{\air}  ]_0    G^{\mp} (u , \chi   ) \kappa^{\mp}_{\flap} (u  , \chi  ) (m_{0 }  \mathbf{u}_0 \cdot \mathbf{U}_0 \mathbf{U}_0  -  \mur  \mathbf{u}_0 \cdot \mathbf{u} \mathbf{u}  )   \left(  \exp\left( 
  \pm \beta_{\air}  \Delta \mu   
 \right)  -  1  \right)   ,
         \end{split}  
 \end{align} 
 where we used the expansion notation
 \begin{align}
       \mathbf{S} =   \mathbf{S}^{[1]}  + \mathcal{O} (\mathbf{u}_0^2 ) ~~ , 
 \end{align}
 and where we introduced the object
 \begin{align} \label{eq:importantffterm321}
   [ f^{(0)}_{\bird}  f^{(0)}_{\air}  ]_0  =    n_{\bird } n_{\air } \left(  \frac{\sqrt{m_{\bird } m_{\air } } }{2 \pi \kb  T_{\air  } } \right)^{3}   \exp( - \beta_{\air }  \left( \frac{1}{2}  (m_0  U_0^2   +  \mur   u^{ 2 }  )   \right) ) ~~ . 
 \end{align}
Upon integrating \eqref{eq:intermediatestep11} over $\mathbf{U}_0$, we find
   \begin{align}  \label{eq:intermediatestep123321}
      \begin{split}
  \mathbf{S}^{[1]}   & =   \frac{K \beta_\air    \mur }{  2 (2 \pi)^2  }    \sum_{\pm} \rho^{\mp} \iint     d    \mathbf{u}  d \mep       \,   u      \mathbf u   G^{\mp} (u , \chi   ) \kappa^{\mp}_{\flap} (u  , \chi  )    
 (   ( \mathbf{u}^{ \pm  \prime }  -  \mathbf{u} )   \cdot \mathbf{u}_0 )  
\exp(    \pm \beta_{\air}  \Delta \mu       )        \exp( -\frac{ \mur   u^{ 2 }   }{2 \kb T_{\air}  }     )  \\ 
   & -    \frac{  K }{ 2 (2 \pi)^2}     \sum_{\pm}  \rho^{\mp}    \iint  d  \mathbf{u}   d \mep   \,   u       G^{\mp} (u , \chi   ) \kappa^{\mp}_{\flap} (u  , \chi  ) (   \mathbf{u}_0    -  \mur  \beta_\air   \mathbf{u}_0 \cdot \mathbf{u} \mathbf{u}  )   \left(  \exp\left( 
    \pm \beta_{\air}  \Delta \mu \right)  -  1  \right)      \exp( -\frac{ \mur   u^{ 2 }   }{2 \kb T_{\air}  }     )   ~~  , 
         \end{split}  
 \end{align} 
 where 
 \begin{align}
     K = \frac{   ( 2  \pi)^2  d^2_{\bird , \air }  n_{\air } }{2}  \frac{  m_\air   }{m_0  }    \left(  \frac{  \mur  \beta_{\air } }{2 \pi } \right)^{\frac{3}{2}}  ~~. 
 \end{align}
 We then integrate over $\me$ as well as the azimuthal relative angle $\epsilon$ to find
  \begin{align}  \label{eq:intermediatestep111}
      \begin{split}
  &  \mathbf{S}^{[1]}    = \\  & 
  \frac{ 1 }{3 }  \beta_\air  \mur      K \mathbf{u}_0   \sum_{\pm} \rho^{\mp}   \iint    d u    d \chi \sin(\chi )       \,   u^5     G^{\mp} (u , \chi   ) \kappa^{\mp}_{\flap} (u  , \chi  )   \left(    \frac{ \sqrt{u^{2} \mp  a^2  }   }{u }     \cos(\chi ) - 1  \right)     
\exp(    \pm \beta_{\air}  \Delta \mu       )        \exp( -\frac{ \mur   u^{ 2 }   }{2 \kb T_{\air}  }     )  \\ 
   & -     K        \mathbf{u}_0  \sum_{\pm}  \rho^{\mp}    \iint  d  u   d \chi \sin(\chi ) \,   u^3        G^{\mp} (u , \chi   ) \kappa^{\mp}_{\flap} (u  , \chi  ) \left(    1   -  \frac{1}{3 }  \beta_\air \mur    u^2 \right)   \left(  \exp\left( 
    \pm \beta_{\air}  \Delta \mu         
 \right)  -  1  \right)      \exp( -\frac{ \mur   u^{ 2 }   }{2 \kb T_{\air}  }     )     , 
         \end{split}  
 \end{align} 
 where we used that
 \begin{align}
\iint d \mathbf{u} \mathbf{e}'  u  \mathbf{u} \mathbf{u}  =  \frac{2 (2 \pi)^2}{3}  \mathbf{1}   \iint d u  d \chi \sin(\chi )  u^5   ~~ , ~~ \iint d \mathbf{u} \mathbf{e}'  u  \mathbf{u} \mathbf{u}^{\prime \pm }  =    \frac{2 (2 \pi)^2}{3}  \mathbf{1}   \iint d u  d \chi \sin(\chi ) \cos(\chi ) u^4 \sqrt{u^2 \mp a^2 }    ~~ , 
 \end{align}
 where $\mathbf{1}$ is the Kronecker delta. Using the definition of \eqref{eq:alphacoefficient}, we can turn \eqref{eq:intermediatestep111} into
     \begin{align}
     \begin{split} \label{eq:flapping321}
               \alpha    =  &  -  \frac{ 1  }{3 }  \beta_\air   \mur   K \sum_{\pm} \rho^{\mp}   \iint    d u    d \chi \sin(\chi )       \,   u^5     G^{\mp} (u , \chi   ) \kappa^{\mp}_{\flap} (u  , \chi  )     \left(     \frac{ \sqrt{u^{2} \mp  a^2  }   }{u }     \cos(\chi ) - 1  \right)     
\exp(    \pm \beta_{\air}  \Delta \mu       )        \exp( -\frac{   \mur   u^{ 2 }   }{2 \kb T_{\air}  }     )  \\  
   &  +        K  \sum_{\pm}  \rho^{\mp}    \iint  d  u   d \chi \sin(\chi ) \,   u^3        G^{\mp} (u , \chi   ) \kappa^{\mp}_{\flap} (u  , \chi  ) \left( 1     -  \frac{1}{3 }  \beta_{\air } \mu_r  u^2 \right)   \left(  \exp\left( \pm \beta_{\air}  \Delta \mu   \right)  -  1  \right)      \exp( -\frac{ \mur   u^{ 2 }   }{2 \kb T_{\air}  }     )   ~~   .       \end{split}
 \end{align}
 Simplifying, we have
      \begin{align}
     \begin{split} \label{eq:flapping12}
               \alpha    =  &  -  \frac{ 1  }{3 }   \beta_\air  \mur   K \sum_{\pm} \rho^{\mp}   \iint    d u    d \chi \sin(\chi )       \,   u^5     G^{\mp} (u , \chi   ) \kappa^{\mp}_{\flap} (u  , \chi  )     \left(    \frac{ \sqrt{u^{2} \mp  a^2  }   }{u }     \cos(\chi )  \exp(    \pm \beta_{\air}  \Delta \mu       )  - 1  \right)     
       \exp( -\frac{   \mur   u^{ 2 }   }{2 \kb T_{\air}  }     )  \\  
   &  +      K  \sum_{\pm}  \rho^{\mp}    \iint  d  u   d \chi \sin(\chi ) \,   u^3        G^{\mp} (u , \chi   ) \kappa^{\mp}_{\flap} (u  , \chi  )      \left(  \exp\left( 
    \pm \beta_{\air}  \Delta \mu         
 \right)  -  1  \right)      \exp( -\frac{ \mur   u^{ 2 }   }{2 \kb T_{\air}  }     )   ~~   .       \end{split}
 \end{align}
\section{Analyzing integrand of $\alpha$}
\label{app:analyze}
\begin{figure*}[!t]
    \centering
   \begin{tikzpicture}
  \node[anchor=south west,inner sep=0] (img) at (0,0)
    {\includegraphics[width=0.45\linewidth]{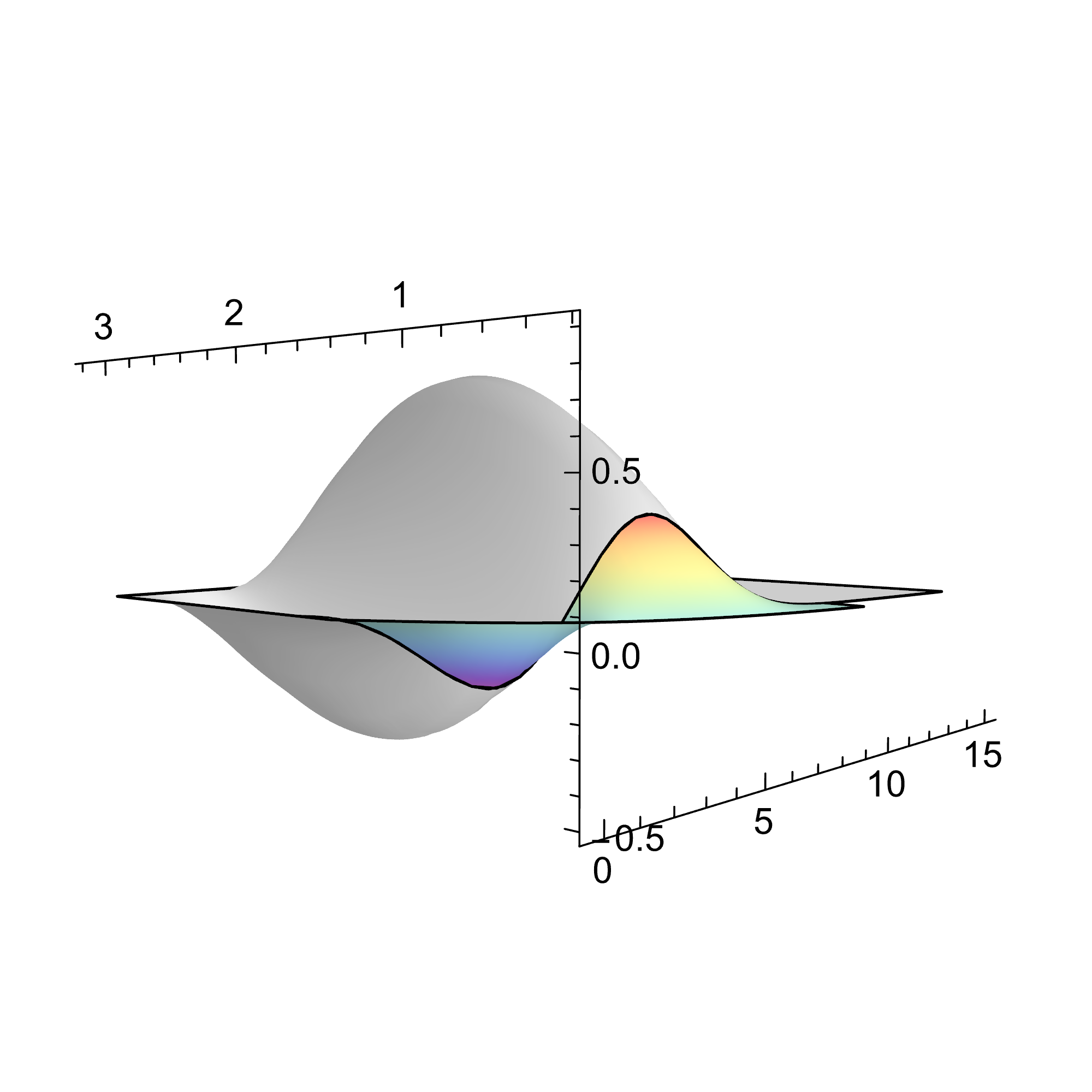}};
  \begin{scope}[x={(img.south east)},y={(img.north west)}]
    \node[fill=white] at (0,0.8) {$\text{(a)}$};
     \node[fill=white] at (0.8,0.2) {$u$};
    \node at (0.3,0.77) {$\chi$};
        \node at (0.43,0.5) {$I_+(u,\chi )$};

  \end{scope}
\end{tikzpicture}
    \hfill
\begin{tikzpicture}
  \node[anchor=south west,inner sep=0] (img) at (0,0)
    {\includegraphics[width=0.45\linewidth]{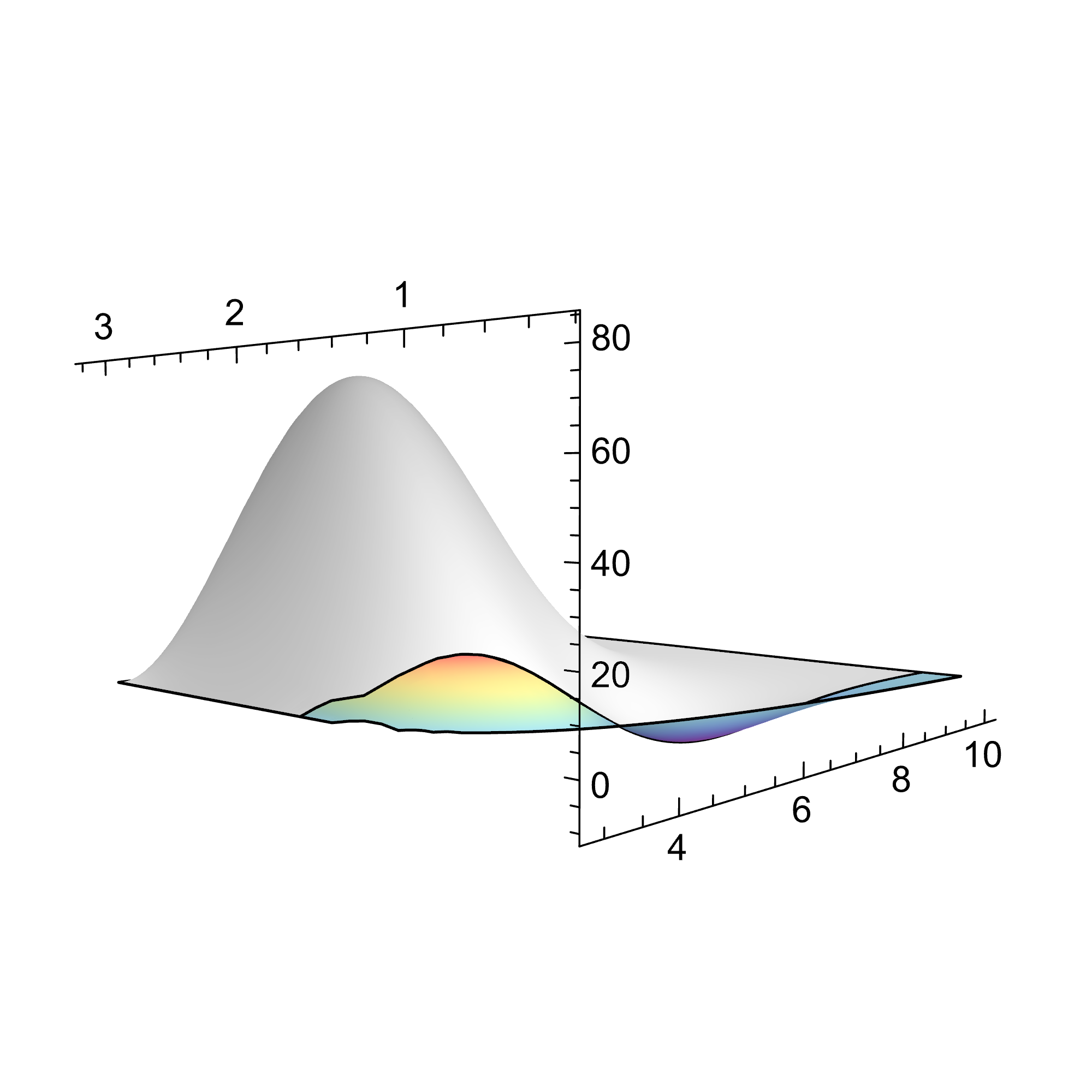}};
  \begin{scope}[x={(img.south east)},y={(img.north west)}]
    \node[fill=white] at (0,0.8) {$\text{(b)}$};
     \node[fill=white] at (0.8,0.2) {$u$};
    \node at (0.3,0.77) {$\chi$};
        \node at (0.43,0.5) {$I_-(u,\chi )$};

  \end{scope}
\end{tikzpicture}
    \caption{The two integrand contributions of \eqref{eq:flapping123}. The gray volume represents the part of the integrand that is cut out by taking $\chi_{\graze}  $ less than $\pi$. We took the same parameter values as in Fig.~\ref{fig:three-subfigures}.}
    \label{fig:twoimages}
\end{figure*}

\textcolor{black}{In this Appendix we analyze the integrand of $\alpha$ given in \eqref{eq:flapping12} with the reactive volume specified by \eqref{eq:rhorho} and \eqref{eq:choice}. Let us split it as
       \begin{align}
     \begin{split} \label{eq:flapping123}
               \alpha    =  & \sum_{\pm}   \iint    d u    d \chi   I_{\pm} ( u , \chi )   ~~   ,        \end{split}
 \end{align}
 where $\pm$ represents the parts that involve $ \kappa^{\pm}_{\flap} ( u , \chi ) $ respectively. In Fig.~\ref{fig:twoimages}. we plot $I_{\pm} ( u , \chi )$ for a particular value of $\chi_{\graze}$, and see that this coefficient has to be small in order for $\alpha$ to be negative and that for most values of $\chi_{\graze}$ $\alpha$ is strongly positive, which is because frontal collisions strongly undermine flocking.}

 \end{document}